\documentclass[
    aps,
    prx,
    letterpaper,
    reprint,
    nobalancelastpage,
    superscriptaddress,
    nofootinbib,
    longbibliography
]{revtex4-2}

\usepackage[colorlinks, citecolor=blue]{hyperref}
\usepackage{graphicx}
\usepackage{amsmath}
\usepackage{amssymb}
\usepackage[english]{babel}
\usepackage{color}
\usepackage[version=4]{mhchem}
\usepackage{dsfont}
\usepackage{multirow}
\usepackage{booktabs}
\usepackage{array}
\usepackage{stmaryrd}
\usepackage{physics}
\usepackage{bm}
\usepackage{comment}
\usepackage{lineno}

\usepackage{soul}
\newcommand{\cP}{ {\cal P} }
\newcommand{\eqnref}[1]{Eq.\,\eqref{#1}}

\newcommand{\figref}[1]{Fig.\,\ref{#1}}

\renewcommand{\t}[1]{\text{#1}}

\newcommand{\UIUC}{
    Department of Physics,
    The University of Illinois at Urbana-Champaign,
    Urbana, IL 61801, USA
}

\renewcommand{\cite}[1]{\mbox{\citep{#1}}}

\addto\captionsenglish{}
\addto\captionsenglish{}

\begin{document}
%\linenumbers

\title{An architecture for two-qubit encoding in neutral ytterbium-171 atoms}
\author{Zhubing Jia}
\affiliation{\UIUC}
\author{William Huie}
\affiliation{\UIUC}
\author{Lintao Li}
\affiliation{\UIUC}
\author{Won Kyu Calvin Sun}
\affiliation{\UIUC}
\author{Xiye Hu}
\affiliation{\UIUC}
\author{Aakash}
\affiliation{\UIUC}
\author{Healey Kogan}
\affiliation{\UIUC}
\author{Abhishek Karve}
\affiliation{\UIUC}
\author{Jong Yeon Lee}\email{jongyeon@illinois.edu}
\affiliation{\UIUC}
\affiliation{Department of Physics, The University of California, Berkeley, CA 94618, USA}
\author{Jacob P. Covey}\email{jcovey@illinois.edu}
\affiliation{\UIUC}

\begin{abstract}
We present an architecture for encoding two qubits within the optical ``clock" transition and nuclear spin-1/2 degree of freedom of neutral ytterbium-171 atoms. Inspired by recent high-fidelity control of all pairs of states within this four-dimensional ``ququart'' space, we present a toolbox for intra-ququart (single-atom) one- and two-qubit gates, inter-ququart (two-atom) Rydberg-based two- and four-qubit gates, and quantum nondemolition (QND) readout. We then use this toolbox to demonstrate the advantages of the ququart encoding for entanglement distillation and quantum error correction which exhibit superior hardware efficiency and better performance in some cases since fewer two-atom operations are required. Finally, leveraging single-state QND readout in our ququart encoding, we present a unique approach to studying interactive circuits and to realizing a symmetry protected topological phase of a spin-1 chain with a shallow, constant-depth circuit.
% These applications are all within reach of recent experiments with neutral ytterbium-171 atom arrays.
\end{abstract}
\maketitle

\section{Introduction}\label{Intro}
Although most quantum computing architectures focus on two-level quantum systems for encoding qubits, the need for auxiliary quantum systems or extra quantum states is ubiquitous. These extra degrees of freedom are used to, e.g., mediate gates between qubits, perform measurements or cooling during a computation, or improve the qubit connectivity. For neutral atom and trapped ion quantum computers, these auxiliary degrees of freedom often take the form of a second atomic species~\cite{Liebfried2003,Singh2021,Singh2023} and/or atoms that are moved during the circuit~\cite{Liebfried2003,Beugnon2007,Lengwenus2010,Pino2021,Bluvstein2022}. Recently, the potential of extra states within the atoms to address these needs has gained intense interest. The ability to use two or more portions of the atomic level structure to perform disparate functions simultaneously without crosstalk has recently enabled ``mid-circuit" operations such as readout and cooling~\cite{Liebfried2003,Monz2016,Yang2022,Graham2023,Lis2023,Ma2023,Scholl2023,Scholl2023b}. Additionally, it has been shown how the ability to programmably repurpose an atom for these roles by moving it between the different sets of levels opens the door to improved hardware efficiency and more flexible circuit compiling~\cite{Allcock2021,Yang2022,Lis2023}.

\begin{figure}[t!]
     \centering
     \includegraphics[width=\linewidth]{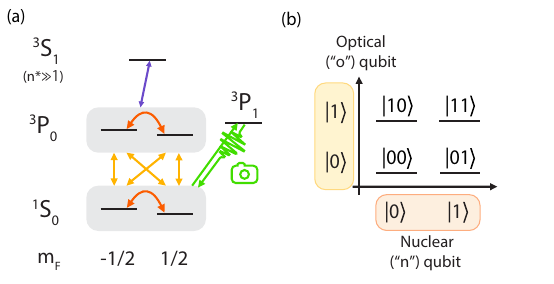}
        \caption{ {\bf Overview of ququart encoding architecture in the omg level structure of $^{171}$Yb atoms}. (a) The relevant energy levels and operations. We encode a ququart in the $^1S_0$ and $^3P_0$ manifolds of $^{171}$Yb atom, each containing two nuclear Zeeman states. Qubit rotations of the states in the same manifold are performed with radio-frequency magnetic field or stimulated Raman transitions, while $^1S_0$ and $^3P_0$ states are connected through optical clock transitions. Readout is performed on $^1S_0$ substates using $^1S_0\leftrightarrow^3P_1$ transition. Inter-ququart entangling gates are driven by single-photon Rydberg excitation from $^3P_0$ states to high-principal quantum number Rydberg states denoted as $^3S_1$. (b) 
        Two-qubit encoding in one ququart. The ququart can be treated naturally as the tensor product of an optical (``o'') and a nuclear (``n'') qubit.
        }
        \label{Fig1}
\end{figure}

When using auxiliary states within an atom to perform ancillary functions for our qubits, a natural question is whether it may be advantageous to include these states in the computational space and encode quantum information in a higher dimension. For instance, when using a pair of extra states to provide two distinct qubit encodings within one atom, we could alternatively think of this four-level system as a ``ququart" composed of two qubits~\cite{Campbell2022}. Higher-dimensional encodings with ``qudits" have attracted much attention for neutral atoms~\cite{Gorshkov2009,Anderson2015,Shi2021a,Omanakuttan2021,Omanakuttan2023,Scholl2023b}, trapped ions~\cite{Low2020, Campbell2022, Ringbauer2022, Low2023, Hrmo2023,Zalivako2024} and superconducting circuits~\cite{Steinmetz2022, Goss2022, Goss2023, Fischer2023}. Qudit systems offer improved hardware efficiency including logical encoding~\cite{Albert2020} and potentially have better performance if the intra-atom gates have higher fidelity than the inter-atom gates. However, the required intra-atom operations quickly become numerous and challenging as the dimension grows, and thus relatively small internal spaces whose states are already widely used for myriad qubit operations offer an attractive starting point for architectures with higher-dimensional encoding.

Here, we focus on a four-level ququart encoding of quantum information in neutral ytterbium-171 ($^{171}$Yb) by employing the optical ``clock" transition and the nuclear spin-1/2 degree of freedom. We show that this ququart can equivalently be represented as two qubits -- the optical (``o") qubit and the nuclear (``n") qubit -- with the full gate set of intra-atom one- and two-qubit gates, inter-atom two- and four-qubit gates, and measurements. We show several example applications of our encoding such as entanglement distillation, quantum error correction, and a new approach to interactive quantum circuits that leverages single-state quantum nondemolition readout. Our work builds upon the growing literature of higher-dimensional encoding schemes for neutral atoms~\cite{Gorshkov2009,Shi2021a,Omanakuttan2021,Scholl2023b}, trapped ions~\cite{Low2020, Campbell2022, Ringbauer2022, Low2023, Hrmo2023} and superconducting circuits~\cite{Steinmetz2022, Goss2022, Goss2023, Fischer2023}, and presents a comprehensive blueprint for qudit encodings that are specifically focused on a direct mapping with qubit circuits~\cite{Campbell2022}.

\section{Results}
\subsection{Ququart operations in $^{171}$Yb atoms}
\label{operations}
We now focus on the level structure and operations surrounding the ququart encoding. Specifically, we focus on operations that are easily translated to the more widely used language of qubits because they allow us to apply our ququart system to any qubit-based algorithm. 
Figure~\ref{Fig1}(a) shows the relevant energy levels and the transitions related to our ququart operations. The ququart is encoded in the $F=1/2{~}^1S_0$ (ground) and $6s6p{~}^3P_0$ (metastable) manifolds, each with two nuclear Zeeman substates. Transitions between the ground and metastable states are driven by a direct optical ``clock" coupling, and transitions between nuclear Zeeman states are driven by a stimulated Raman coupling. The mapping between our ququart states and two-qubit states ($\ket{00}$, $\ket{01}$, $\ket{10}$ and $\ket{11}$) is based on treating the ququart states as the tensor product of an optical clock (``o''-) qubit and a nuclear Zeeman (``n''-) qubit, as is shown in Fig.~\ref{Fig1}(b). We discuss this mapping more formally and consider ququart state tomography in the Supplementary Materials (SM)~\cite{SuppMat}.

Before going into detail on this encoding scheme and the required operations, we briefly summarize the state of the art and the primary challenges. The ground and metastable nuclear spin qubits have both been manipulated with gate fidelities of $\mathcal{F}\approx0.999$ and have coherence times of $T_2^*>7$ seconds~\cite{Barnes2021,Ma2022,Jenkins2022,Huie2023,Norcia2023,Ma2023}. The primary challenge centers around realizing fast, high-fidelity optical qubit operations with a long coherence time. Long-term phase stability is needed for running long circuits, high-frequency phase noise must be suppressed when performing fast gates, and deleterious motional effects from finite temperature and finite trap frequencies must be mitigated. Techniques have been devised to filter high-frequency phase noise on ultrastable lasers~\cite{Endo2018,Li2022,Lis2023,Chao2023}, and motional ground-state cooling schemes~\cite{Jenkins2022,Lis2023,Scholl2023b} combined with carefully chosen pulse design~\cite{Scholl2023b,Lis2023} have been used to obviate the effects of atomic motion. Finally, we assume that the atoms are trapped at a ``magic" wavelength for which the potential experienced for all four states is identical~\cite{Ye2008,Huie2023,Lis2023,Norcia2023}. Optical clock transitions have been manipulated with gate fidelities of $\mathcal{F}\approx0.998$ and have coherence times of $T_2^*\approx3$ seconds~\cite{Norcia2019,Lis2023,Scholl2023b}.

\begin{figure}[t!]
    \centering
    \includegraphics[width=\linewidth]{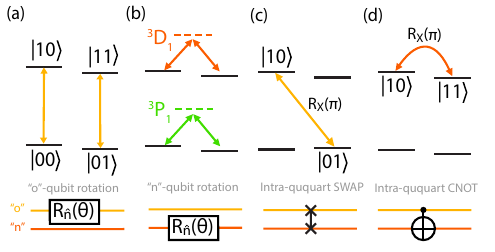}
    \caption{ 
    {\bf Intra-ququart gates through clock and Raman transitions}. (a) Single-qubit rotations $R_{\hat{n}}(\theta)$ of angle $\theta$ along rotation axis $\hat{n}$ on the ``o''-qubit via pairs of clock transitions. To ensure that the state of the other qubit is unaffected, the two transitions within a pair should perform the same rotations on each pair of states. (b) Single-qubit rotations $R_{\hat{n}}(\theta)$ on the ``n''-qubit driven by pairs of stimulated Raman transitions. (c) The intra-ququart swap gate is achieved by applying a $\pi$-rotation between $\ket{01}$ and $\ket{10}$ states. This operation swaps the information encoded in ``o''- and ``n''-qubits. (d) Intra-ququart CNOT gate is achieved by applying a $\pi$-rotation between $\ket{10}$ and $\ket{11}$ states. When the control qubit (``o''-qubit) is in $\ket{1}$ state, a bit flip operation is applied to the target qubit (``n''-qubit).}
    \label{Fig2}
\end{figure}

\subsubsection{Intra-ququart gates}
\label{1qgate}

Using the two-qubit encoding described above, intra-ququart one- and two-qubit gates can be performed by driving optical clock or Raman transitions. Figure~\ref{Fig2} shows four typical sets of intra-ququart gates, including single-qubit rotations, two-qubit swap, and CNOT gates. Further details on the intra-ququart gates can be found in the SM~\cite{SuppMat}. 

The single-qubit gates on the ``o''- or ``n''-qubit require one pair of clock or Raman transitions mediated by different states [Fig.~\ref{Fig2}(a)(b)]. For the ``o''-sector rotations, we assume that both transitions have the same resonance frequency. This is a good approximation in a small magnetic field~\cite{Lis2023} or under application of a light shift on one of the four states when operating at a high field (see Ref.~\cite{SuppMat}). However, even when this condition is not satisfied and the two sectors undergo a slow, passive entangling operation, this effect can be deterministically compensated. For the ``n''-qubit rotations, the Raman transitions in the ground and metastable states do not have crosstalk since the wavelengths associated with the intermediate states are separated by $\sim 10^2$ THz. The Rabi frequencies of each Raman transition can be matched.

Intra-ququart two-qubit gates are often simpler than single-qubit gates: Figure~\ref{Fig2}(c)(d) shows the intra-ququart swap and CNOT operations. The swap gate is achieved by driving a $\pi$-pulse between $\ket{01}$ and $\ket{10}$ state, effectively swapping the information encoded in ``o''-qubit and ``n''-qubit. The CNOT gate with the control qubit ``o'' and the target qubit ``n'' is achieved by swapping the population in $\ket{10}$ and $\ket{11}$ states. 
Note that operations such as the CNOT gate induce a differential light shift in the `o' sector, leading to a phase accumulation during its execution. This effect can be neutralized with an additional light shift or deliberately engineered to produce a trivial $2\pi$ phase shift.

\begin{figure}[t!]
    \centering
    \includegraphics[width=\linewidth]{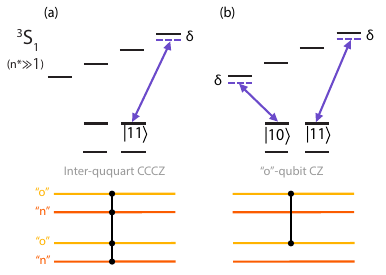}
    \caption{ {\bf Inter-ququart gates based on Rydberg transitions}. Rydberg coupling is accessed through a single-photon transition of $\lambda=302$ nm from $^3P_0$ states to high-principal quantum number states. (a) One stretched Rydberg transition is driven on two nearby atoms, achieving inter-atom entanglement via Rydberg blockade. With the two-qubit encoding, this inter-ququart gate performs a CCCZ gate. (b) Inter-ququart CZ gates on the ``o''-qubits of two ququarts. Both stretched Rydberg transitions are driven with the same Rabi frequency, detuning, and phase such that both $^3P_0$ states can couple to Rydberg states. }
    \label{Fig3}
\end{figure}

\subsubsection{Inter-ququart gates}
\label{section:2qgate}

For neutral atoms, entangling two physical qubits is commonly achieved through Rydberg blockade~\cite{Isenhower2010, Saffman2010, Wilk2010, Jau2016, Graham2019, Levine2019}. Specifically, two-qubit entangling gates between metastable nuclear spin qubits in $^{171}$Yb~\cite{Ma2023} and optical clock qubits in strontium-88~\cite{schine2022,Scholl2023b} have been achieved with single-photon Rydberg excitations. To perform an entangling gate, one commonly used way is to selectively couple one of the two qubit states to a Rydberg state, such that due to Rydberg blockade between nearby atoms, the qubit states accumulate a state-dependent phase, generating a CZ gate under proper experimental settings~\cite{Levine2019}. 

With our ququart encoding, we first consider a four-qubit CCCZ gate: the CZ gate that was recently demonstrated for the metastable nuclear spin qubit of $^{171}$Yb~\cite{Ma2023} becomes a CCCZ gate directly with the inclusion of the two ground states into the computational space. As shown in Fig.~\ref{Fig3}(a), the phase shift due to Rydberg interactions is only present in the $|11\rangle$ state of each ququart.The gate's performance is simulated in~\cite{SuppMat}, where we demonstrate a theoretical maximum effectiveness at the 99.9$\%$ fidelity level. 

We also demonstrate an inter-ququart CZ gate between the two ``o''-qubits of the ququarts. Figure~\ref{Fig3}(b) shows the gate scheme of this ``o"-sector inter-ququart CZ gate, which must be agnostic to the ``n" sector. To excite \textit{both} the states in the optical $\ket{1}$ state to the Rydberg manifold, we drive two stretched Rydberg transitions between $6s6p{~}^3P_0\ket{m_F=\pm 1/2}$ and $6sns{~}^3S_1\ket{m_F=\pm 3/2}$ states with the same Rabi frequency, detuning, and phase. 
Apart from the necessity of these dual-tone Rydberg excitation, our approach adheres to the standard Rydberg CZ gate protocol~\cite{Levine2019, Jandura2022}. This gate scheme requires Rydberg states of different $m_F$'s to blockade each other. In~\cite{SuppMat}, we show that the blockade strengths between Rydberg states of different $m_F$'s are expected to be of the same order of magnitude (limited spectroscopic data is available), and we perform a numerical simulation of this ($2\times8$)-dimensional space for realistic parameters. We believe that, by adding a sideband to the Rydberg laser pulse, it is relatively straightforward to achieve this CZ ququart gate with a performance matching that of the `standard' two-qubit CZ gate.

Notably, we can also perform gates between an atom with ququart encoding and an atom functioning only as a qubit which is encoded in the metastable nuclear spin. There are two approaches:  One option is to apply only one Rydberg drive ($\sigma^+$ \textit{or} $\sigma^-$) to the qubit atom but apply both ($\sigma^+$ \textit{and} $\sigma^-$) to the ququart atom. This will perform a CZ gate between the metastable nuclear spin of the qubit atom and the ``o''-qubit of the ququart atom. A second option is to first convert the metastable nuclear spin of the qubit atom into an optical qubit via the swap gate in Fig.~\ref{Fig2}(c), then apply the CZ gate with both Rydberg tones on each atom (global pulses), and then swap the qubit-atom's encoding back to the metastable nuclear spin.

\subsubsection{Readout}
\label{readout}
The readout of higher-dimensional spaces like qudits can be cumbersome. In neutral atoms and trapped ions, the ground state manifold typically exhibits fluorescence (``bright"), in contrast to the non-fluorescent (``dark") metastable manifolds. The usual method for qudit readout in such cases involves 'shelving' all but one basis state into the metastable manifold, with the chosen state transitioned (mapped) into the ground state for readout. Following fluorescence readout, this ground state is reset and then shelved into the metastable manifold, with the process sequentially repeated for each state~\cite{Low2023}. Hence, readout of a $d$-level qudit usually involves up to $d-1$ iterations of this cycle as only one state is probed at a time. This approach is exponentially less efficient than measuring multiple qubits in parallel.

\begin{figure}[t]
     \centering
     \includegraphics[width=\linewidth]{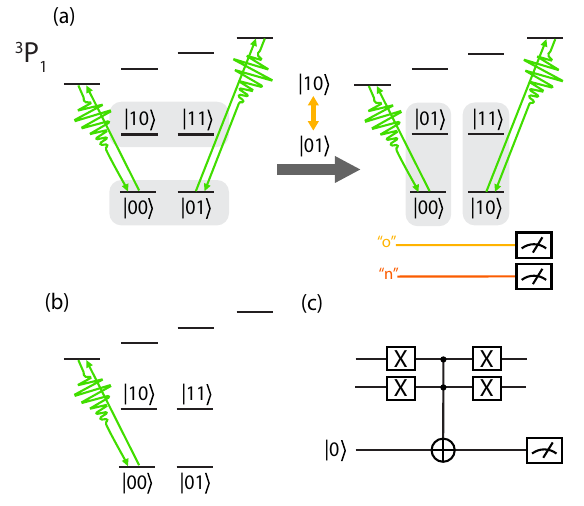}
        \caption{{\bf Different ququart readout schemes}. (a) 2-round deterministic readout of ququart states. 
        Two stretched transitions $^1S_0\leftrightarrow$~$^3P_1$ are simultaneously driven, and the resultant fluorescences are detected in an indistinguishable manner, effectively measuring the ``o''-qubit.
        Note that the cyclicity of the transitions ensures that the population in $^1S_0$ substates remains in the same state after measurement. Afterward, an intra-ququart SWAP is applied, swapping the populations in $\ket{01}$ and $\ket{10}$ states. Finally, the same two stretched transitions are used to effectively measure the ``n''-qubit before the swap operation. The four combinations of two readout results can thus be mapped to the four ququart states. (b) Single-state QND readout of $\ket{00}$, within the $^1S_0$ ground state manifold. (c) The two-qubit analog of the single-state QND measurement via an ancilla qubit and a Toffoli gate.}
        \label{Fig4}
\end{figure}

Instead, here we perform a 2-round deterministic readout of the ququart states based on the quantum nondemolition (QND) readout of the $^1S_0$ nuclear spin ground states~\cite{Huie2023, Norcia2023}, which reduces the upper limit of the required number of measurement operations from 3 to 2. Figure~\ref{Fig4}(a) shows the 2-round ququart readout scheme. This readout protocol requires simultaneously probing on two $^1S_0\leftrightarrow$~$^3P_1$ stretched transitions and collecting the scattered photons. Both stretched transitions are closed (``cycling") under a magnetic field of $\gtrsim10^2$ G, which allows us to perform readout of the ``o''-qubit without transferring the population between ``n" qubit states. We apply two rounds of the dual-stretched readout with an intra-qubit SWAP operation sandwiched in between. Conceptually, the first readout measures the ``o''-qubit while the second readout measures the ``n''-qubit. The two bits of information from the readout process determine the projective measurement outcome and post-measurement state. For instance, ``bright"-``bright" refers to $\ket{00}$ while ``bright"-``dark" refers to $\ket{01}$, etc. 

Note that readout of only one qubit within the ququart without affecting the other qubit remains challenging with this readout scheme. One exception is that if the ``o''-qubit QND readout measures $\ket{1}$, then the ``n''-qubit is encoded only in the $^3P_0$ states, which is left unaffected. Therefore, the information encoded in the ``n" qubit is preserved. As described below, this condition is sufficient for an entanglement distillation protocol in which ``o"-qubit readout with outcome $\ket{1}$ heralds the generation of purified entanglement in the ``n''-sector. Accordingly, although an ``o"-qubit readout outcome of $\ket{0}$ would collapse the ``n''-qubit, the distillation protocols fail anyway and must be re-attempted. We also note that by separately detecting the fluoresced light from the left- and right-stretched transitions using polarization or frequency filtering, it would be possible to fully determine the ququart state in one measurement round when the ``o"-qubit is in $|0\rangle$.

Finally, we consider a unique opportunity for qudit readout. The typical shelving-probing protocol discussed above that requires $d-1$ rounds is cumbersome if all basis states must be measured. However, this scenario presents a unique opportunity to perform a QND readout of a portion of the computational space. Specifically, we propose to perform single-state QND readout via probing with only one stretched transition from the ground manifold~\cite{Huie2023, Norcia2023} [see Fig.~\ref{Fig4}(b)]. In this case, $\ket{00}$ will be ``bright" while all other ququart states will be ``dark". A measurement outcome of ``bright" will collapse the ququart to $\ket{00}$, while an outcome of ``dark" allows a re-normalized superposition of $\ket{01}$, $\ket{10}$, and $\ket{11}$ to survive. Naturally, we must be aware of light shifts on the $\ket{01}$ state due to the probe light as discussed in \cite{SuppMat}. The two-qubit analog of this single-state QND readout paradigm requires an ancilla qubit and a non-Clifford gate as shown in Fig.~\ref{Fig4}(c). As shown below, this paradigm provides new opportunities for interactive circuits.

\subsection{Applications}
\label{Applications}

Qudit architectures offer improved hardware efficiency, enabling the encoding of more quantum information -- potentially including logical qubits -- within a single atom. Moreover, the fidelity of two-atom gates often still lags behind that of one-atom gates in both trapped ion and neutral atom systems. Consequently, the proposed ququart architecture promises potential performance enhancements over traditional qubit systems. This potential is contingent upon maintaining the quality of one-atom gates amidst the complexities introduced by the ququart's expanded gate set and its additional constraints. However, we note that a portion of the required gate set falls within the toolbox of the `omg' qubit-based architecture. Moreover, the requirements for light shifts and local control are not enormously beyond those required for qubit-based architectures. As such, we believe that intra- and inter-ququart operations could match the performance of `standard' intra- and inter-qubit operations, for which single-atom gates still outperform two-atom gates.

In this section, we focus on leveraging this improved hardware efficiency, 
enhanced effective connectivity, and potentially superior performance, as well as on applications that exploit the unique readout properties of qudit systems for interactive circuits. In the discussions that follow, we use black lines in the circuit diagrams to represent standard qubits, while two adjacent lines, colored yellow and orange, denote the ``o"- and ``n"-qubits of a ququart, respectively.

\subsection{Quantum information processing}
\subsubsection{Entanglement distillation}

Entanglement distillation is a protocol to generate a high-fidelity entangled two-qubit state from two low-fidelity entangled two-qubit states using local operations and classical communication~\cite{Bennett1996, Bennett1996a}, which has been demonstrated in several physical platforms with qubit architectures~\cite{Kwiat2001, Pan2003, Kalb2017}.

Within our proposed ququart architecture, the distillation protocol is executed more efficiently, requiring only two atoms with operations internal to the ququart and a global readout mechanism. This contrasts with traditional qubit systems, where the distillation necessitates four atoms, the application of inter-atomic gates, and mid-circuit (local) readout techniques.

\begin{figure}[t!]
     \centering
     \includegraphics[width=\linewidth]{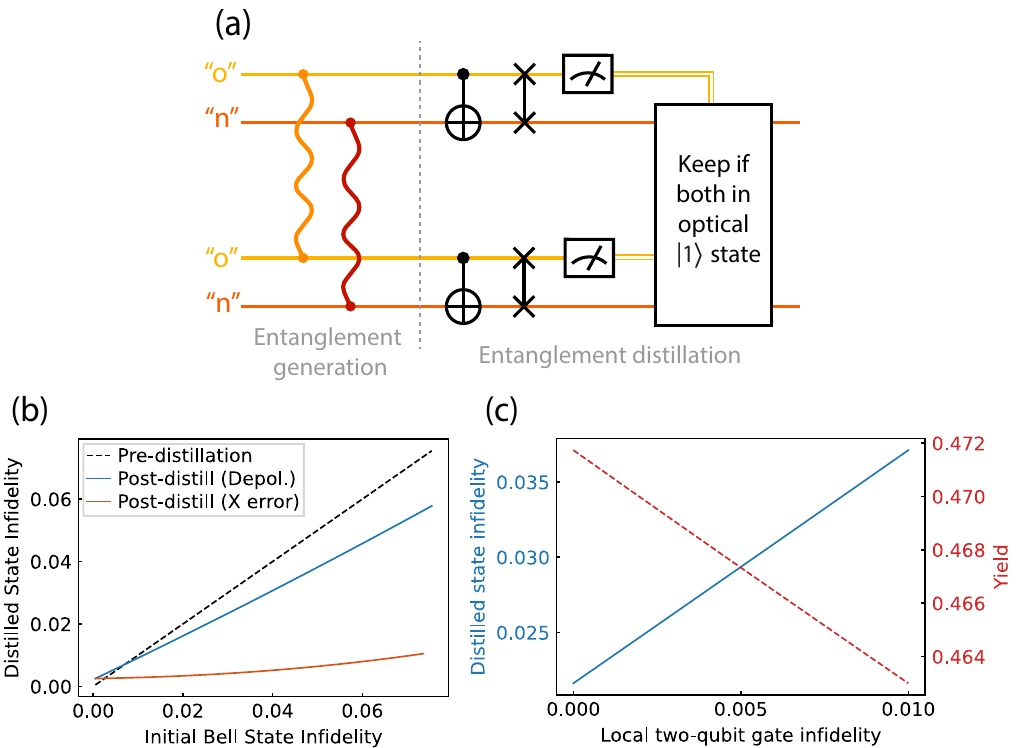}
        \caption{{\bf Entanglement distillation with two ququarts}. 
        (a) Circuit diagram of entanglement distillation, with two nearby yellow and red qubit as a ququart. The two curved lines denote entanglement between two qubits in two ququarts. After the entanglement generation, intra-ququart two-qubit gates are applied, followed by a measurement on the ``o''-qubits. We only keep the post-distilled state when both ``o''-qubits measure $\ket1$, which indicates that the entanglement purification succeeds and the ``n''-qubit coherence is unaffected. 
        (b) Simulation results of Bell state infidelities before and after distillation assuming depolarization and Pauli-X errors.  
        (c) Simulation results of distilled state infidelity and yield under different local two-qubit gate infidelities. Here the pre-distilled state is generated through a depolarization channel with pre-distilled state infidelity of 3\%.
        }
        \label{Fig5}
\end{figure}

As illustrated in Figure~\ref{Fig5}(a), the entanglement distillation protocol begins with a pair of two-qubit entangled states. In our ququart architecture, we assume that the pre-distilled state is given as follows:
\begin{align}
    |\Psi_0 \rangle = \bigotimes_{ i \in \{\textrm{o}, \textrm{n}\} } 
    \frac{1}{\sqrt{2}}\big( \, |0_{1} \rangle^i |0_{2} \rangle^i  + |1_{1} \rangle^i |1_{2} \rangle^i \,\big).
\end{align}
Here, the subscript indicates the atom, while the superscript distinguishes whether the qubit is optical (o) or nuclear (n). We highlight the presence of entanglement between the optical qubits across two distinct atoms, as well as between their nuclear qubits. This arrangement, comprising a pair of Bell states, can be efficiently produced using inter-ququart gates.

We want the final two-qubit state to be encoded in the nuclear qubits of two atoms. Hence, we apply SWAP operations to both ququarts so that we preserve ``n" qubits rather than ``o" qubits. Finally, the ``o" qubit in each ququart is measured in the Pauli $Z$-basis and the success of the distillation protocol is heralded by an outcome of $\ket{1}$ (dark) for both, such that the resulting two-qubit state is indeed encoded in nuclear qubits. We emphasize that this ``o''-qubit readout is performed globally as the optical qubit state $|1 \rangle$ can be measured without interfering with nuclear qubits as discussed in Sec.~\ref{readout}.

In Figure~\ref{Fig5}(b), we present the infidelities between both pre-distilled and distilled states compared to the ideal Bell state, aiming to showcase the circuit's efficacy in distillation. To this end, we consider two distinct error models impacting the inter-ququart gates during the entanglement generation step: the depolarization channel and Pauli-X noise. For both scenarios, we factor in a consistent measurement error rate of 1\% and an intra-ququart gate infidelity of 0.1\%, with both types of errors simulated via the depolarization channel. The yield under these two types of errors is shown in Fig. \ref{Fig:Yield} in Methods.

The distillation circuit is specifically designed to mitigate Pauli-$X$ errors on Bell states. However, it does not address Pauli-$Z$ errors, leading to its diminished effectiveness in combating depolarization errors, as illustrated by the plot. Given the varied nature of inter-ququart gate errors, which realistically may fall between these two models, we anticipate an improvement in Bell state fidelity under certain conditions. Naturally, higher fidelity local operations in the distillation protocol will lead to higher yield and lower distilled state infidelity. In Fig.~\ref{Fig5}(c), we plot both these metrics as a function of local two-qubit gate infidelity for a 3\% depolarization infidelity on the pre-distilled Bell pairs. Under a realistic assumption that our intra-ququart two-qubit gates have higher fidelity than inter-atom two-qubit gates, our ququart distillation protocol can significantly outperform the standard, qubit-based approach.
Consequently, the post-distillation states are expected to provide high-fidelity entangled qubit pairs, proving beneficial for a range of applications involving metastable qubits.

\subsubsection{Flag-based quantum error correction}
\label{flag}

For large-scale quantum computing, quantum error correction (QEC) becomes necessary to overcome hardware imperfections~\cite{Shor1995, Gottesman1997, Preskill1998, Terhal2015}. QEC is achieved by encoding logical qubits using multiple physical data qubits. With this redundancy, one can measure specific multi-body operators called \emph{stabilizers} to identify errors in the physical qubits without interfering with logical qubits, thereby determining the necessary corrections. When the error rate of the physical qubits and their associated operations are below a specific threshold, logical qubits can achieve an error rate lower than the base error rate of the physical qubits. A QEC code that employs $n$ physical qubits to encode $k$ logical qubits with code distance $d$ is denoted as a $\llbracket n,k,d\rrbracket$ code.

However, QEC is effective only if the quantum circuits that identify and correct errors do not proliferate those errors across physical qubits -- a requirement called fault-tolerance (FT)~\cite{Gottesman1998, gottesman2009introduction}. Unfortunately, a straightforward approach using a single ancilla qubit (syndrome qubit) for stabilizer measurement generally fails to meet this FT standard. To address this, various strategies have been proposed, including the use of a block of ancilla qubits~\cite{Shor1996, Steane1997, Knill2005}.

More recently, a novel approach emerged to minimize the large number of ancilla qubits traditionally needed. By introducing one extra \emph{flag} qubit in stabilizer measurements, one can track and contain an error that could have proliferated through the circuit~\cite{Chao2018, Chamberland2018, Chao2020} with relatively low qubit overhead (only two auxiliary qubits are required). This concept has shown promising results in solid-state spin qubits~\cite{Abobeih2022}.

\begin{figure}[t!]
     \centering
     \includegraphics[width=\linewidth]{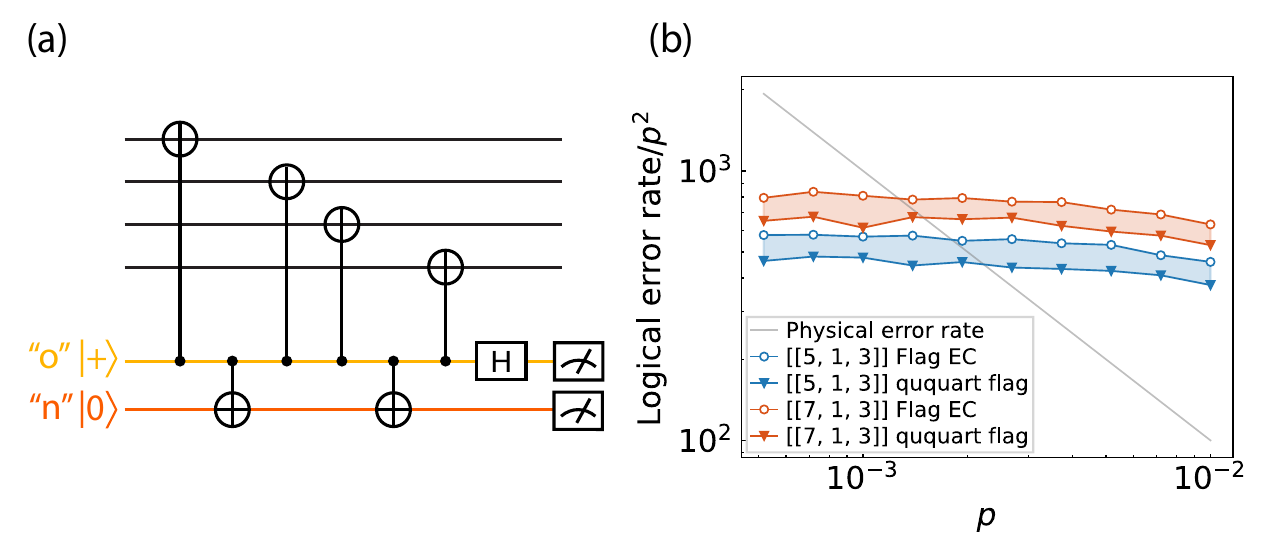}
        \caption{ {\bf Flag-based QEC with ququarts} (a) Circuit to fault-tolerantly extract the stabilizer $XXXX$. 
        % using the two qubits encoded in a ququart as syndrome and flag qubits. 
        Black lines are data qubits and yellow and red lines are syndrome and flag qubits encoded in a ququart. The CNOTs between syndrome and flag qubits are performed with intra-ququart gates, which should suppress the infidelity to the same level as single-qubit gates. (b)
        Simulated threshold for $\llbracket5,1,3\rrbracket$ and Steane $\llbracket7,1,3\rrbracket$ codes. Circles represent normal flag EC and triangles represent flag EC with ququart by assuming intra-ququart gates have perfect fidelity. The intersection of the grey line (physical error rate) and the curve representing logical error rate of a certain case gives the physical error rate, below which the QEC code outperforms the bare qubit. Using ququart flag improves the threshold, and in practice, the threshold should lie in between the normal flag case and the ququart flag case.}
        \label{Fig6}
\end{figure}

With ququarts, this flag-based QEC can be naturally implemented as illustrated in Fig.~\ref{Fig6}(a). The ``o''-qubit serves as the ancilla qubit while the ``n''-qubit serves as the flag qubit. Two-qubit gates between a normal qubit and the ``o''-qubit within a ququart can be performed directly if the qubit is encoded in the optical transition, or by swapping its nuclear spin encoding to the optical sector before and after the gate (see section~\ref{section:2qgate}). With high-fidelity intra-ququart two-qubit gates between the ancilla and flag qubits, an increase in the error correction performance is possible. Additionally, using just a single auxiliary atom for fault-tolerant operations reduces the number of atoms and shuttling operations that are required to achieve a specified level of qubit connectivity, thereby reducing the hardware overhead.

To demonstrate the advantage of having ququarts in flag-based QEC, we calculated logical error rates with gates of an error rate $p$ for two different scenarios in Figure~\ref{Fig6}(b): In the first scenario, the error rate between flag and syndrome qubits is also $p$, while in the second scenario, the error rate is zero. The latter case corresponds to the ququart implementation with an almost-perfect intra-ququart gate. In practice, the intra-ququart gates generally have higher fidelity than the gates between ququart and qubit, so the ququart flag performance should lie in the shaded region in Fig.~\ref{Fig6}(b). We compared the performance using two types of QEC code: the five-qubit $\llbracket5,1,3\rrbracket$ code and Steane seven-qubit $\llbracket7,1,3\rrbracket$ code. Details about the simulation can be found in Methods. For both QEC codes, using a ququart flag increases the error threshold due to the higher-fidelity intra-ququart gates. 

\subsubsection{Four-qubit code error detection}

Another useful direction towards fault-tolerant quantum computation is the quantum error detection code. One example of quantum error detection code is the smallest instance of surface codes, the four-qubit $\llbracket4, 2, 2\rrbracket$ code~\cite{Vaidman1996}. Unlike QEC codes that can correct errors on physical qubits, the four-qubit code can detect errors on a single qubit but is not able to correct the errors. The four-qubit code can act as a building block for large-scale fault-tolerant quantum computations~\cite{Knill2005}, and fault-tolerant operations on the four-qubit code have been demonstrated in several physical platforms~\cite{Linke2017, Andersen2020, Takita2017}. 

The four-qubit code is defined by the stabilizers
\begin{align}
\begin{split}
    S_X &= X_1X_2X_3X_4,\\
    S_Z &= Z_1Z_2Z_3Z_4.
\end{split}
\end{align}
Here $X_i$ and $Z_i$ are Pauli X and Z operators on the $i$-th physical qubit. The logical operators assigned to two logical qubits are
\begin{align}
\begin{split}
    X_{L1} &= X_1X_3,~Z_{L1} = Z_1Z_2,\\
    X_{L2} &= X_1X_2,~Z_{L2} = Z_1Z_3.    
\end{split}
\end{align}
The logical states are
\begin{align}
\begin{split}
    \ket{00}_L &= \left(\ket{0000}+\ket{1111}\right)/\sqrt{2},\\
    \ket{01}_L &= \left(\ket{0011}+\ket{1100}\right)/\sqrt{2},\\
    \ket{10}_L &= \left(\ket{0101}+\ket{1010}\right)/\sqrt{2},\\
    \ket{11}_L &= \left(\ket{0110}+\ket{1001}\right)/\sqrt{2}.
\end{split}
\end{align}
Specifically, one can fix the first logical qubit as a gauge and the code turns into a $\llbracket4,1,2\rrbracket$ subsystem code.

\begin{figure}[t!]
     \centering
     \includegraphics[width=\linewidth]{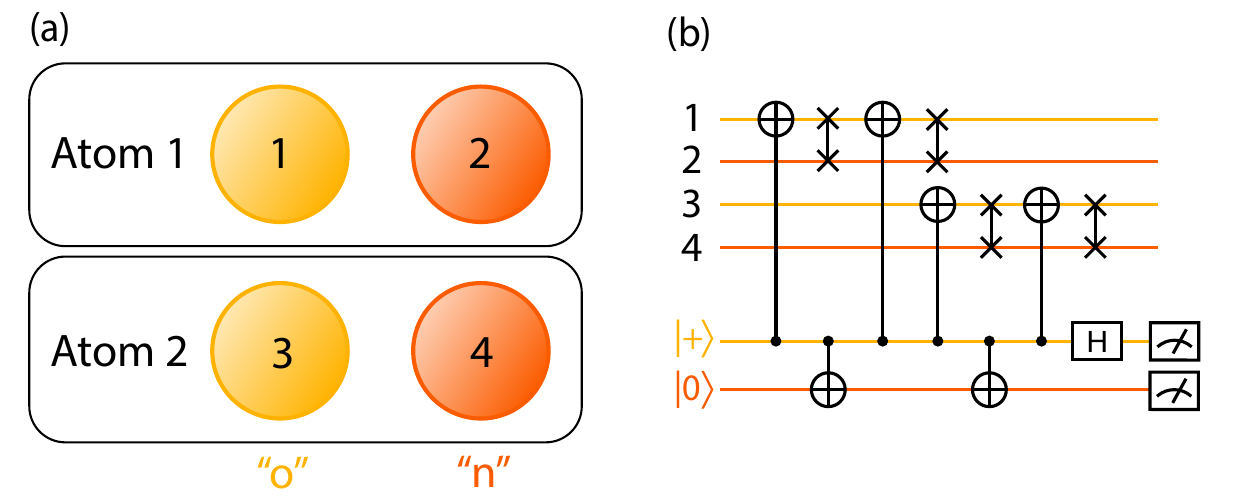}
        \caption{ {\bf Four-qubit code encoded in two ququarts}. (a) The ``o''-qubit and ``n''-qubit of two ququarts are used to encode the four-qubit code. (b) Measurement of the stabilizer $S_{X}$ using an extra ancilla ququart. The extra ququart flag is implemented to fault-tolerantly measure the stabilizer without introducing logical errors into the state.}
        \label{Fig7}
\end{figure}

We encode the four-qubit code using the ``o'' and ``n'' qubits of two ququarts, as shown in fig.~\ref{Fig7}(a). To avoid correlated two-qubit errors from stabilizer measurement, an auxiliary ququart is implemented to function as the ququart flag discussed in section~\ref{flag} (see fig.~\ref{Fig7}(b)). Therefore, the fault-tolerant operations on the four-qubit code can be demonstrated using only three individual atoms. We further highlight that the specific logical states,
\begin{align}
    \ket{0\pm} = \left(\ket{00}\pm\ket{11}\right)^{\otimes 2}/2,
\end{align}
can be easily prepared with two ququarts by applying clock $\pi/2$-rotations between the $\ket{00}$ and $\ket{11}$ states. Additionally, the use of fewer atoms in the ququart protocol increases the effective connectivity such that the need for coherent transport is obviated.

\subsection{Interactive circuits}
\subsubsection{Realizing topological phases}

\begin{figure*}[t]    
\includegraphics[width=\textwidth]{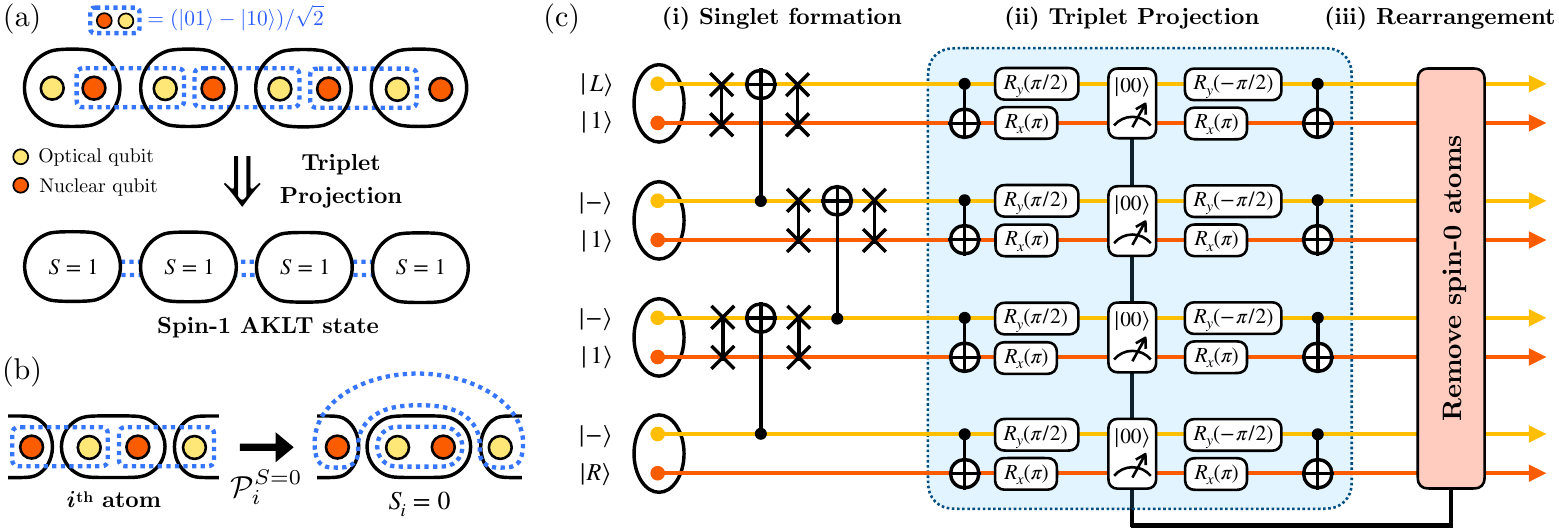}
\caption{    \label{fig:AKLTprotocol}
 {\bf AKLT state preparation protocol} {\bf(a)} Within each ququart, there are two qubits (optical and nuclear). Starting from a chain of singlets between optical and nuclear qubits across different sites, we project each ququart onto the spin-1 subspace. The projected wavefunction is the exact ground state of the AKLT Hamiltonian.
 {\bf(b)} If $i$-th ququart is projected onto the spin-0 subspace instead, the ``o''-qubit of the $(i\,{-}\,1)$-th ququart and ``n''-qubit of the $(i\,{+}\,1)$-th ququart automatically forms a maximally entangled singlet. {\bf(c)} The quantum circuit of the actual experimental procedure consists of three steps. 
 ($i$) Initialize the system into the product state of a particular pattern, which is followed by the intra-ququart SWAP and inter-ququart CNOT to create a chain of singlets.  
 $(ii)$ Apply individual qubit rotations and intra-ququart gates to rotate $|01 \rangle - |10 \rangle$ into $|00 \rangle$, measure $|00 \rangle$, and then apply the reverse operation. This sequence effectively performs singlet/triplet projections.
 $(iii)$ After the measurement, ququarts projected onto the spin-1 subspace automatically form the AKLT state, while spin-0 ququarts completely disentangle. One can rearrange neutral atoms to throw away spin-0 ququarts.}
\end{figure*}

One fascinating aspect of ququarts is that a natural measurement does not bisect the local Hilbert space unlike in a system made up of qubits~\footnote{For a quantum circuit consisting of qubits, such a measurement operation would require introducing an extra ancilla qubit and applying a non-Clifford gate.}. The quantum channel for measuring $| 00 \rangle$ in a ququart is expressed as:
\begin{align} \label{eq:measureChannel}
  &  \rho \otimes |0 \rangle \langle 0|_E  ~ \mapsto ~ \cP_0 \rho \cP_0 \otimes |0\rangle \langle 0|_E + \cP_{\bar{0}} \rho \cP_{\bar{0}} \otimes |\bar{0}\rangle \langle \bar{0}|_E  \nonumber \\
    &\quad  \cP_0 = |0 0 \rangle \langle 0 0 | \nonumber \\
    &\quad  \cP_{\bar{0}} =  |0 1 \rangle \langle 0 1| + |1 0 \rangle \langle 1 0| + |1 1 \rangle \langle 1 1|,
\end{align}
where $|0 \rangle_E$ and $|1 \rangle_E$ are the states of a measurement apparatus. 
Using any intra-qubit unitary transformation $U$ rotating $|00 \rangle$ into $(|10\rangle - |01 \rangle)/\sqrt{2}$, through this measurement one can distinguish spin-singlet ($s=0$) and spin-triplet ($s=1$) sectors of a given ququart. 
This leads to a new protocol to prepare interesting quantum many-body wavefunctions.

To illustrate our approach, consider a spin-1 chain model by Affleck, Kennedy, Lieb, and Tasaki (AKLT)~\cite{AKLT1987, AKLT1988} defined by the following Hamiltonian:
\begin{align}\label{eq:AKLT}
    H_\textrm{AKLT} = \sum_i \Big[ (\vec{\bm{S}}_i \cdot \vec{\bm{S}}_{i+1}) + \frac{1}{3} (\vec{\bm{S}}_i \cdot \vec{\bm{S}}_{i+1})^2 \Big],
\end{align}
where $\vec{\bm{S}}_i = (S^x_i, S^y_i, S^z_i)$ is the spin-1 operator at site $i$.
Its ground state is a canonical example of nontrivial symmetry protected topological (SPT) order, exhibiting several interesting properties~\cite{AKLT0, AKLTstringorder, AKLT1}: ($i$) \emph{Non-local order parameter:} For any arbitrary length $|i-j|$, the expectation value of the following \emph{string} of operators takes a finite value:
\begin{align}
    O^\alpha_{ij} := {S}^\alpha_i e^{i \pi \sum_{l=i+1}^{j-1} {S}^\alpha_l} {S}^\alpha_j, \qquad \alpha=x,y,z.
\end{align}
The finite expectation value of a nonlocal operator $O^\alpha_{ij}$ implies that the state can be utilized as an entanglement resource for quantum computation~\cite{AKLTQC_1,AKLTQC_2,AKLTQC_3,AKLTQC_4}. ($ii$) \emph{Anomalous boundary modes:} In the open boundary condition, effective spin-1/2 edge degrees of freedom exist at each edge albeit consisting of spin-1 in the microscopic model.

There are various protocols suggested to realize this state using \emph{qubits} and local unitary gates. However, these schemes require a gate depth linearly scaling~\cite{AKLT_exp0, AKLT_exp1, AKLT_exp2, AKLT_exp3, AKLT_exp4, AKLT_exp5, AKLT_exp6} with the system size due to a finite correlation length of the AKLT state, which requires some exponential tail of unitary gates. As we will show, this can be circumvented by our novel approach.

A schematic diagram for our protocol is shown in \figref{fig:AKLTprotocol}(a). First, we prepare a wavefunction where an ``o''-qubit at the $i^\mathrm{th}$ site and a ``n''-qubit at the $(i{+}1)^\mathrm{th}$ site form a singlet $\frac{1}{\sqrt{2}}(|0\rangle_i^\textrm{o} |1\rangle_{i+1}^\textrm{n}\,{-}\,|1\rangle_i^\textrm{o} |0\rangle_{i+1}^\textrm{n})$. Second, we project each ququart onto its spin-triplet ($S\,{=}\,1$) manifold via measurement. If all ququarts are properly projected, they immediately form the AKLT state, which is the ground state of \eqnref{eq:AKLT}. However, each ququart Hilbert space decomposes into spin-singlet and triplet sectors, and the probability of getting a spin-0 measurement outcome is $25\%$. Thus, it is natural to ask whether we can get a whole chain of the AKLT state instead of a set of small AKLT islands. 

The solution to this problem is straightforward: by discarding neutral atoms that are projected onto spin-singlets, we can ensure that the remaining ququarts will form a single AKLT state. To elucidate this idea, let us examine the scenario illustrated in \figref{fig:AKLTprotocol}(b), where we label qubits as 1, 2, 3, and 4 in a cardinal manner. Initially, the qubits (1,2) and (3,4) are in singlet states, resulting in a zero total spin. Now, we project the pair $(2,3)$ onto a singlet. Given that the initial total spin is zero and that the conservation of total spin is a requisite in each projected component, it logically follows that the pair (1,4) must also be in a spin-singlet state. As a result, while ququarts projected onto spin-singlets become disentangled from the rest of the system, those projected onto spin-triplets contribute to forming a single AKLT state. Therefore, our approach is distinguished from other proposals, such as Ref.~\onlinecite{AKLT_exp7}.

Using experimental controls discussed in the previous section, the protocol consisting of three parts can be directly implemented as shown in \figref{fig:AKLTprotocol}(c):
\begin{itemize}
    \item {\bf Inter-ququart Singlet Formation}: As our inter-ququart operations are limited among optical qubits, we have to introduce SWAP operations. Using intra-ququart SWAP and inter-ququart CNOT operations in a parallel manner, this step can be done within five steps.
    \item {\bf Triplet Projection}: To perform this projection, we have to conjugate our $|00 \rangle$ measurement by the basis rotation using single-qubit rotations and intra-ququart CNOT gates. 
    \item {\bf Rearrangement}: Conditioned on the measurement outcomes, we can remove atoms projected onto spin-singlet states either virtually (through post-processing) or physically (through the optical tweezer technique~\cite{Bluvstein2022}). 
\end{itemize}
The size of this resultant AKLT state is approximately 3/4 of the original system, and most importantly, there is no exponential overhead in this protocol.

Although the proposed method is inherently stochastic, we can also create an AKLT state of a specific size through an additional step. 
The key idea is that any two-qubit state can be projected into a singlet by Bell state measurement up to a Pauli operator, and this Pauli operator can be corrected by a local unitary that depends on the measurement outcome. 
In achieving a target size, there are two approaches, bottom-up and top-down. In the bottom-up approach, we use a \emph{fusion} protocol to merge two different AKLT segments. In the top-down approach, we use a \emph{fission} protocol to remove sites from a single AKLT segment. The details of fusion and fission protocols are elaborated in Methods.

Therefore, our protocol combines local unitary gates, measurements, and feedback to create an AKLT state in a constant depth, offering a novel experimental route to realize exotic topological systems in quantum simulator platforms. We note that the proposed method can be readily extended for various one and two-dimensional fixed-point SPT states where symmetry actions on bulk degrees of freedom can be pushed to fractional degrees of freedom at its boundaries~\cite{_ahino_lu_2021, PhysRevB.94.205150, Molnar_2018, RevModPhys.93.045003}. 

Interestingly, the illustrated protocol can be considered as a particular experimental implementation of a so-called parton construction, which is heavily utilized in the theoretical study of quantum many-body physics. In this construction, an exotic quantum state is created by starting from a simple wavefunction consisting of \emph{fractional} degrees of freedom and then projecting it onto physical degrees of freedom~\cite{PartonArovasAuerbach1988, PartonJain1989, PartonReadSachdev1991, PartonWen2002, PartonWen2002_2}. Therefore, our approach paves the road to implement this parton construction in experiments.

\subsubsection{Adaptive dynamics: absorbing transitions}
\label{MIPT}

The capacity to perform mid-circuit measurements and feedback operations based on the outcomes in ququarts offers experimental access to \emph{adaptive dynamics}, where the system evolves under random local unitaries, measurement, and local feedback operations based on measurement outcomes. In recent years, the entanglement structure of such adaptive dynamics has been theoretically explored where the measurement probability $p$ of each qubit for each cycle of the circuit has been the main tuning knob. 
In particular, two interesting transitions have been theoretically discussed: the volume to area-law entanglement transition across a critical measurement rate~\cite{Skinner2019, Li2018}, and directed percolation or steering transitions between two distinct area law phases across another critical tuning parameter~\cite{piroli2023, odea2022entanglement, Sierant2023}.

Our ququart encoding, or more generically higher-dimensional encodings offer a unique opportunity to explore these phenomena. In fact, the qudit single-state QND readout discussed above has been utilized in recent theoretical proposals for studying directed percolation transitions~\cite{piroli2023}. 
% In this context, we will define our ququart state generically by $|\psi\rangle=\alpha|0\rangle+\beta|1\rangle+\gamma|2\rangle+\delta|3\rangle$. 
Our single-state QND readout of $|00\rangle$ makes it an ``absorbing state": A ququart that is measured to be in this state becomes ``inactive'' and does not participate in subsequent unitary operations. Conversely, if the ququart is measured to be \textit{not} in $|00\rangle$, then it remains ``active'' and its Hilbert space dimension reduces from four to three. Accordingly, above a certain measurement rate, the system transitions into a many-body absorbing state whose universality class is given as a directed percolation transition.

\begin{comment}
\begin{figure}[t]
     \centering
     \includegraphics[width=\linewidth]{MIPTs.pdf}
        \caption{{\color{blue} It is unclear whether this figure is really necessary as it is simple reproduction of other paper; } Directed percolation transitions with ququart circuits. (a) Each ququart is initially in $|\psi\rangle=\alpha|0\rangle+\beta|1\rangle+\gamma|2\rangle+\delta|3\rangle$. We then apply a layer of brickwork two-ququart gates. This is followed by single-ququart reset to the ``inactive" state $|0\rangle$ operations randomly applied with probability $p$. Finally, the $|0\rangle$-state QND readout is applied to all ququarts which either collapses them to $|0\rangle$ (``inactive") or leaves them in a generic three-dimensional space spanned by $|1\rangle$, $|2\rangle$, and $|3\rangle$ (``active"). (b) Maybe a plot here?}
        \label{Fig8}
\end{figure}
\end{comment}

We now outline an exemplary circuit to study measurement-induced phase transitions (MIPTs) with qudits as introduced in Ref.~\cite{piroli2023}. We perform a $|00\rangle$-state measurement of all ququarts in each cycle. After the first measurement layer, the active and inactive sectors remain separated. The knob that we will use to re-couple the sectors and introduce phase transitions are ``reset'' operations in which select ququarts are re-initialized to $|00\rangle$, which can be achieved via local optical pumping. Note that single-ququart dissipative operations can be performed without deleterious effects on the other ququarts by trapping the target ququart(s) in tweezers of a different clock-magic wavelength than the other ququarts for which the probe/reset transition is sufficiently light-shifted from that of the nominal tweezers. Mid-circuit readout has already been demonstrated via this technique~\cite{Norcia2023}. The probability $p$ of performing such a reset operation on each ququart in each cycle is the parameter that allows us to traverse the phase diagram.

This architecture presents a relatively simple approach to studying MIPTs, specifically directed percolation transitions. Since it is known to be very challenging to study entanglement transitions between volume and area law phases in large systems due to the exponential post-selection problem~\cite{Skinner2019, Li2018}, the study of percolation transitions between area law phases appears to be a better target for large-scale quantum hardware~\cite{piroli2023, odea2022entanglement, Sierant2023}. Hence, a concrete and feasible route towards this realization constitutes an important development for interactive quantum circuits.

\section{Discussion}
We have presented a toolbox for encoding two qubits within the four-level structure of alkaline earth-like atom isotopes with nuclear spin-1/2 like $^{171}$Yb. We focused on the application of this ququart encoding to entanglement distillation and quantum error correction. We have also discussed how the higher-dimensional encoding offers new opportunities for interactive circuits to prepare topological phases and study measurement-induced phase transitions. This ququart architecture is a natural extension of the `omg' architecture~\cite{Allcock2021,Chen2022,Lis2023} in which all pairs of states within the four-level system are used to encode qubits with various functionalities. We believe the use of ququarts does not require further capabilities than the `omg' scheme yet offers a significant hardware advantage in terms of atom count and effective connectivity as well as new opportunities. We believe that the performance of our ququart operations can potentially match those of qubit-based operations, and we note that many of our required operations are also required in the `omg' architecture. Additionally, we anticipate that single-atom gates will continue to outperform two-atom gates, and thus larger code spaces can offer improved circuit performance. We note that while the use of the optical qubit is perhaps the largest challenge in our protocol, it could be possible to instead use larger nuclear spin manifolds in, e.g., $^{173}$Yb and $^{87}$Sr to realize a similar multi-qubit architecture.

Alkaline earth-like atoms without nuclear spin may still provide access to our ququart encoding. For instance, we could replace the nuclear spin-1/2 degree of freedom with two harmonic motional states~\cite{Scholl2023b}. Motional ground state preparation via Raman sideband cooling ~\cite{Jenkins2022,Lis2023} or optically-resolved sideband cooling techniques~\cite{Cooper2018,Norcia2018} including erasure cooling~\cite{Scholl2023b} have provided access to temperatures for which this encoding is realistic. Indeed, there is at least one scenario in which the motional encoding may offer an advantage: A crucial capability that is proposed in this work is the inter-ququart gate that performs a CZ gate between the optical qubit (``o'') sector of each ququart without affecting the nuclear qubit (``n'') sector. If we instead encode the second qubit in the motional sector and do not have nuclear spin, then this inter-ququart gate between  ``o''-sectors would be given for free with a single-tone Rydberg excitation. In other words, the standard Rydberg-mediated gate is already agnostic to the motional state in the low temperature regime.

\section*{Methods}

\subsection{Two-ququart Rydberg-based gates}
%\subsection{Analyzing the inter-ququart gate protocol}
We propose two flavors of inter-ququart gates using the single-photon Rydberg coupling from the ${}^3{P}_0$ nuclear spin states. Both gate protocols extend that detailed in Ref.~\cite{Levine2019} to the two-ququart computational space. Operating on the space of states $\{\ket{00} \otimes \ket{00},\, \ket{00} \otimes \ket{01},\, \ket{00} \otimes \ket{10},\, \dots, \ket{11} \otimes \ket{11}\}$, the first is a CCCZ gate that applies a phase shift $\varphi$ to all states for which exactly one atom is in the $\ket{11}$ ququart state, a different phase shift $\varphi'$ to the $\ket{11} \otimes \ket{11}$ state such that $2 \varphi - \pi = \varphi'$, and zero phase shift to the remaining states. The second gate is a CZ gate between only the ``o'' sectors of the two ququarts, applying $\varphi$ to any states of the forms $\ket{1a} \otimes \ket{0b}$ or $\ket{0a} \otimes \ket{1b}$, $\varphi'$ to any states of the form $\ket{1a} \otimes \ket{1b}$, and zero phase shift to the remaining states.

The CCCZ gate has the simpler of the two implementations, requiring a drive on only a single transition from the ${}^3{P}_0 ~ m_F = +1/2$ clock state to the $n\,{}^3{S}_1 ~ F = 3/2,\, m_F = +3/2$ ($n = 59$, for example) Rydberg state [see Fig.~\ref{Fig3}(a)]. The protocol itself has been experimentally realized using this transition in Ref.~\cite{Ma2022}; we only analyze its effects on the total ququart space. The analysis of this gate is straightforward through numerical simulation (see below).

On the other hand, the CZ gate requires the applied phase shifts to be fully agnostic to the ``n'' sectors of both ququarts, for which the pulse sequence must then be duplicated with identical detuning and power on the transition between the ${}^3{P}_0 ~ m_F = -1/2$ and $n\,{}^3{S}_1 ~ m_F = -3/2$ states [Fig.~\ref{Fig3}(b)]. This second configuration also requires that atoms in states of different $m_F$ in the $n\,{}^3{S}_1$ manifold experience an energy shift that is similar to that for states of identical $m_F$ -- a configuration that has not yet been explicitly studied to our knowledge.

In the sections that follow, we perform numerical simulations of the proposed ququart CZ and CCCZ protocols. In our calculations, we neglect effects from laser phase noise and Doppler shifts due to finite temperature since we seek to only highlight challenges arising from the higher-dimensional encoding. For the ququart CZ gate, we examine interactions between Rydberg states of differing $m_F$ values and use the results in our simulations. 

\subsubsection{Rydberg interactions for multiple Zeeman states}
\label{ZeemanBlockade}
Determination of the $\sigma_z$-type interaction strength between Rydberg states of differing $m_F$ values amounts to a calculation of the $C_6$ van der Waals interaction potential coefficient. We base our calculations on the perturbative treatment of the problem in Ref.~\cite{Vaillant2014} for $I = 0$ bosonic species, which gives the value of the $C_6$ coefficient for different pair states as the eigenvalues of a $\mathcal{C}_6$ matrix whose elements are calculated as
\begin{equation}
    \label{eq:c6mat}
    (\mathcal{C}_6)_{ij}
        = -\sum_p \Bigg[
            \frac{\big|\mathcal{R}_{11}(s_0, s_p)\big|^2}{E_0 - E_p}
            \mathcal{D}_{11}(s_i, s_p, \hat{\rho})
            \mathcal{D}_{11}(s_p, s_j, \hat{\rho})
        \Bigg]
.\end{equation}
This sum runs over $p$ indexing pair states of the form $s_p = (n, \alpha, m_J)$ where $n$ and $m_J$ are the principal and total electronic angular momentum projection quantum numbers and, for brevity, $\alpha$ is a vector containing all other relevant total-spin quantum numbers. $E_p$ is the energy of state $s_p$, and the subscript $0$ denotes the initial eigenstate of the unperturbed Hamiltonian. $\hat{\rho}$ is a unit vector $\hat{\rho} = (\theta, \varphi)$ describing the angles in spherical coordinates between the two atoms relative to the quantization axis.

The functions $\mathcal{R}_{11}$ and $\mathcal{D}_{11}$ are the appropriate radial and angular parts of the matrix elements of the electric dipole operator,
\begin{align}
    \mathcal{R}_{k_1, k_2}&(s', s)
        \nonumber\\
        &\approx \left(
            \int_0^\infty \text{d}r_1\,
                R_{n_1', L_1'}(r_1)
                R_{n_1, L_1}(r_1)
                r_1^{2 + k_1}
        \right)
        \nonumber\\
        &\times \left(
            \int_0^\infty \text{d}r_2\,
                R_{n_2', L_2'}(r_2)
                R_{n_2, L_2}(r_2)
                r_2^{2 + k_2}
        \right)
    \\
    \label{eq:dme-angular}
    \mathcal{D}_{k_1, k_2}&(s', s, \hat{\rho})
        \nonumber\\
        &= (-1)^{k_2}
            \left[
                4 \pi
                \frac{(2 k_1 + 2 k_2)!}{(2 k_1)! (2 k_2)!}
                \frac{(2 L_1 + 1) (2 L_2 + 1)}{(2 k_1 + 2 k_2 + 1)}
            \right]^{1/2}
        \nonumber\\
        &\times \left[ (2 J_1 + 1) (2 J_2 + 1) \right]^{1/2}
            C_{L_1 0, k_1 0}^{L_1' 0} C_{L_2 0, k_2 0}^{L_2' 0}
        \nonumber\\
        &\times \begin{Bmatrix} J_1 & k_1 & J_1' \\ L_1' & S & L_1 \end{Bmatrix}
            \begin{Bmatrix} J_2 & k_2 & J_2' \\ L_2' & S & L_2 \end{Bmatrix}
        \nonumber\\
        &\times
            \sum_{q = -(k_1 + k_2)}^{k_1 + k_2}
            \sum_{q_1 = -k_1}^{k_1}
            \sum_{q_2 = -k_2}^{k_2} \Big[
        \nonumber\\
        &\quad\quad
            C_{k_1 q_1, k_2 q_2}^{(k_1 + k_2) q}
            C_{J_1 m_{J1}, k_1 q_1}^{J_1 m_{J1}}
            C_{J_2 m_{J2}, k_2 q_2}^{J_2 m_{J2}}
            Y_{k_1 + k_2, q}(\hat{\rho}) \Big]
\end{align}
where $R_{n, \ell}(r)$ is a Hydrogen radial wavefunction, $C_{J_1, m_1; J_2, m_2}^{J_3, m_3}$ is the Clebsch-Gordan coefficient $\braket{J_1, m_1; J_2, m_2}{J_3, m_3}$, and $Y_{\ell, m}$ is a spherical harmonic. At the time of writing, the Rydberg states of ${}^{171}\t{Yb}$ are not yet well-studied, and we lack sufficient quantum defect data for a proper calculation of $\mathcal{R}_{11}$. Given the results of Refs.~\cite{Ma2022, Ma2023}, however, we assume a nominal Rydberg interaction following $C_6' = 5\,\text{THz}\,\text{$\mu$m}^6$ for $n = 74$, scaled down to the $n = 59,~ {}^3\t{S}_1 ~ F = 3/2$ Rydberg states, and focus on only the angular factors in \ref{eq:c6mat}. Thus we reduce the calculation to
\begin{equation}
    \label{eq:c6mat-angular}
    (\tilde{\mathcal{C}}_6)_{ij}
        = -\sum_p \Big[
            \mathcal{D}_{11}(s_i, s_p, \hat{\rho})
            \mathcal{D}_{11}(s_p, s_j, \hat{\rho})
        \Big]
\end{equation}
in order to show only that the expected variation in the size of the Rydberg interaction within the targeted manifold would not be disruptive to the implementation of the CZ gate.

We therefore rewrite Eq.~\ref{eq:dme-angular} by way of the Wigner-Eckart theorem as
\begin{widetext}
    \begin{equation}
        \begin{aligned}
            \mathcal{D}_{k_1, k_2} ~ \rightarrow ~ \tilde{\mathcal{D}}_{k_1, k_2}&(s', s, \hat{\rho})
                \\
                &= (-1)^{
                    - F_1 + k_1 - m_{F1}'
                    - F_2 + k_2 - m_{F2}'
                    + J_1' + J_2'
                    - L_1 - L_2
                    + 2 I
                    + 2 S
                    + k_1 + k_2
                }
                \\
                &\times (-1)^{k_2}
                    \left[
                        \frac{4 \pi}{(2 k_1 + 2 k_2 + 1)}
                        \frac{(2 k_1 + 2 k_2)!}{(2 k_1)! (2 k_2)!}
                    \right]^{1/2}
                \\
                &\times \big[
                    (2 F_1' + 1) (2 F_2' + 1)
                    (2 J_1' + 1) (2 J_2' + 1)
                    (2 L_1' + 1) (2 L_2' + 1)
                    \\
                    &\quad\quad \times
                    (2 F_1  + 1) (2 F_2  + 1)
                    (2 J_1  + 1) (2 J_2  + 1)
                    (2 L_1  + 1) (2 L_2  + 1)
                \big]^{1/2}
                \\
                &\times \begin{Bmatrix} J_1' & F_1' & I \\ F_1 & J_1 & k_1 \end{Bmatrix}
                    \begin{Bmatrix} J_2' & F_2' & I \\ F_2 & J_2 & k_2 \end{Bmatrix}
                    \begin{Bmatrix} L_1' & J_1' & S \\ J_1 & L_1 & k_1 \end{Bmatrix}\\
                &\times
                    \begin{Bmatrix} L_2' & J_1' & S \\ J_2 & L_2 & k_2 \end{Bmatrix}
                    \begin{pmatrix} L_1' & k_1 & L_1 \\ 0 & 0 & 0 \end{pmatrix}
                    \begin{pmatrix} L_2' & k_2 & L_2 \\ 0 & 0 & 0 \end{pmatrix}
                \\
                &\times
                    \sum_{q = -(k_1 + k_2)}^{k_1 + k_2}
                    \sum_{q_1 = -k_1}^{k_1}
                    \sum_{q_2 = -k_2}^{k_2}
                        \begin{pmatrix} F_1' & k_1 & F_1 \\ -m_{F1}' & q_1 & m_{F1} \end{pmatrix}
                        \begin{pmatrix} F_2' & k_2 & F_2 \\ -m_{F2}' & q_2 & m_{F2} \end{pmatrix}
                        C_{k_1 q_1, k_2 q_2}^{(k_1 + k_2) q}
                        Y_{(k_1 + k_2) q}(\hat{\rho})
        \end{aligned}
    \end{equation}
\end{widetext}
using the particular choices $S_1 = S_2 = S_1' = S_2' \equiv S = 1$ and $I_1 = I_2 = I_1' = I_2' \equiv I = 1/2$, which are motivated below. Following the diagonalization of Eq.~\ref{eq:c6mat-angular}, we calculate $\tilde{C}_6$ values for all possible pair states in a chosen basis. These values are dimensionless and act only as a weighting factor to estimate the strength of the Rydberg interaction between a given pair of states in the $n\,{}^3\t{S}_1$ manifold. Specifically, the $\tilde{C}_6$ value for the $k$-th pair state is calculated as
\begin{equation}
    \tilde{C}_6^{(k)}
        = \sum_j \braket{k}{j} \bra{j} \tilde{C}_6 \ket{j} \braket{j}{k}
        %= \sum_j \lambda_j |\mathbf{e}_k \cdot \mathbf{v}_j|^2
\end{equation}
where $\ket{j}$ is the $j$-th eigenvector of $\tilde{\mathcal{C}}_6$. The final $C_6$ value for the $k$-th pair state used in numerical simulations (see below) is then $C_6^{(k)} = \big( C_6' / \max_k |\tilde{C}_6^{(k)}| \big) \times \tilde{C}_6^{(k)}$.

Although we only consider these angular weighting factors for interactions between states within the targeted Rydberg manifold, the calculation as a whole still depends critically on effects from other, nearby Rydberg states whose contributions to Eq.~\ref{eq:c6mat} eventually vanish as the energy difference factors in the denominator become large and the overlaps between radial wavefunctions become small. Lacking both in Eq.~\ref{eq:c6mat-angular}, we are then forced to decide independently on a particular basis over which the sum is carried out. From the analysis in previous work~\cite{Chen2022}, we also restrict the basis to states for which $S = 1$ and $F = I + J$. Thus the chosen basis comprises all hyperfine states associated with the manifolds $\{{}^3\t{S}_1\, F = 3/2,~ {}^3\t{P}_0\, F = 1/2,~ {}^3\t{P}_1\, F = 3/2,~ {}^3\t{P}_2\, F = 5/2,~ {}^3\t{D}_1\, F = 3/2, \dots\}$ which we choose to cut off at ${}^3\t{D}_1\, F = 3/2$ (for a total of 20 states) as an estimated threshold for where contributions to Eq.~\ref{eq:c6mat} would mostly vanish in an ideal calculation. Note that Eq.~\ref{eq:c6mat-angular} does not depend on the principal quantum number $n$ because we have discarded all energy and radial factors. The results of the full calculation for the targeted ${}^3\t{S}_1\, F = 3/2$ manifold are shown in Table~\ref{tab:c6-vals}.

\begin{table}[htbp]
    \renewcommand*{\arraystretch}{1.5}
    \centering
    \caption{
        \textbf{$C_6 / C_6'$ weighting factors between ${}^3\t{S}_1\, F = 3/2$ hyperfine states} calculated from Eq.~\ref{eq:c6mat-angular}, normalized to their largest-magnitude value. The row indexes the $m_F$ quantum number of the first atom while the column indexes that of the second. Note that the Rydberg interaction is symmetric with respect to the interchange of the two states.
        \label{tab:c6-vals}
    }
    \begin{tabular}{ccccc}
        \hline\hline
        $m_F^{(1)}$ & \multicolumn{4}{c}{$m_F^{(2)}$} \\
        \cmidrule{1-1} \cmidrule(l){2-5}
               & $-3/2$   & $-1/2$   & $+1/2$   & $+3/2$   \\
        $-3/2$ & $-0.989$ & $-0.980$ & $-0.984$ & $-1.000$ \\
        $-1/2$ & $-0.980$ & $-0.969$ & $-0.970$ & $-0.984$ \\
        $+1/2$ & $-0.984$ & $-0.970$ & $-0.969$ & $-0.980$ \\
        $+3/2$ & $-1.000$ & $-0.984$ & $-0.980$ & $-0.989$ \\
        \hline\hline
    \end{tabular}
\end{table}

\subsubsection{Numerical simulations of Rydberg gates}
We examine both the CCCZ and CZ ququart gate protocols by numerically simulating the ground-clock-Rydberg state manifold using the methods from previous work~\cite{Chen2022} under the application of the pulse sequence described in Ref.~\cite{Levine2019}. The actions and fidelities of the gate protocols were evaluated by simulating the appropriate pulse sequence with the two-ququart system initialized in each computational basis state. The simulations were performed for a typical Rabi frequency $\Omega_\text{Ryd} = 2 \pi \times 3\,\text{MHz}$ on the targeted Rydberg transitions (with laser detuning determined by the Rabi frequency; $\Delta / \Omega_\text{Ryd} \approx 0.375$ for both protocols) and nominal interaction strength $2\,\t{GHz}$ in the presence of a magnetic field $B = 120\,\text{G}$. The authors of Ref.~\cite{Levine2019} cite finite atom temperature and off-resonant scattering as major limitations to the fidelity of their experimental realization. The CZ and CCCZ protocols presented here inherit these limitations; however to examine effects arising purely due to the extension of the original protocol to the ququart space, our simulations are conducted with effects from neither of these sources, nor from e.g. laser phase noise. For both CZ and CCCZ, all excitation pulses were performed for horizontally polarized light with angle of incidence perpendicular to the quantization axis. This adds extra couplings to those shown in Fig.~\ref{Fig3}; namely $\ket{11} \leftrightarrow {}^3\text{S}_1\, \ket{m_F = -1/2}$ for both CCCZ and CZ, and $\ket{10} \leftrightarrow {}^3\text{S}_1\, \ket{m_F = +1/2}$ for CZ. Both are far off resonance, detuned by approximately twice the Zeeman splitting in the Rydberg manifold.

Notably, the forms of both gates are defined in terms of two phases $\varphi$ and $\varphi' = 2 \varphi - \pi$ where $\varphi$ can in general be any real number modulo $2 \pi$, which leaves the ``ideal'' form of the gate (in the case of perfect fidelity) parameterized by $\varphi$. We therefore calculate the fidelity of our simulated gates according to
\begin{align}
    \mathcal{F}_\text{CCCZ}
        &= \max\limits_{\varphi}
            \frac{1}{16} \big|\tr\big(\text{CCCZ}^\dagger(\varphi) ~ \text{CCCZ}'\big)\big|
    \\
    \label{eq:cz-fidelity}
    \mathcal{F}_\text{CZ}
        &= \max\limits_{\varphi}
            \frac{1}{16} \big|\tr\big(\text{CZ}^\dagger(\varphi) ~ \text{CZ}'\big)\big|
\end{align}
where $\text{CCCZ}(\varphi)$ and $\text{CZ}(\varphi)$ are the ideal realizations of the gates, $\text{CCCZ}'$ and $\text{CZ}'$ are the forms of the gates realized from simulation, and we normalize by the identity operator on the total $(4 \times 4)$-state two-ququart basis. It was assumed that the two ground states would not have significant coupling to any other nearby states from the Rydberg excitation pulses, and it was calculated that any associated light shifts from dipole-allowed transitions could be neglected at the $\sim 10^{-5}\,\text{Hz}$ level. Under the conditions identified above, we find $\mathcal{F}_\text{CCCZ} \gtrsim 0.999$ and $\mathcal{F}_\text{CZ} \gtrsim 0.999$. We do not identify additional limitations in $\mathcal{F}_\text{CCCZ}$ associated with the ququart encoding; however, $\mathcal{F}_\text{CZ}$ depends on the assumed Rydberg interactions, as we now describe.

\begin{figure}[t]
    \includegraphics[width=\columnwidth]{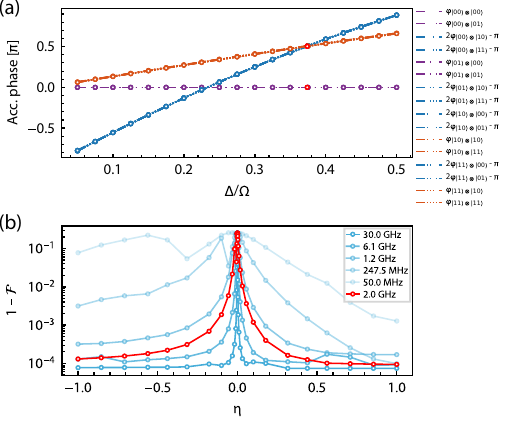}
    \caption{
        \textbf{Inter-ququart CZ gate performance}. (a) Accrued phases $\varphi$ for all two-ququart basis states in units of $\pi$ as a function of $\Delta / \Omega$ for $\eta = 0.989$, corresponding to the weighting values shown in Table~\ref{tab:c6-vals}, and overall Rydberg energy shift $2\,\text{GHz}$. Lines are colored by the number of 1's in the ``o'' positions of both atoms' ququart bitstrings -- 0 (purple), 1 (blue), and 2 (orange), giving distinct accruals $\varphi_0$, $\varphi_1$, and $\varphi_2$, respectively. The dot-dash patterns of each of the lines correspond to their associated states' two-ququart bitstrings. The CZ gate protocol from Ref.~\cite{Levine2019} is implemented for $2 \varphi_1 - \pi = \varphi_2$ (red points). (b) CZ gate infidelities as a function of $\eta \in [-1, 1]$, parameterizing the uniformity of the Rydberg interaction between the various states of the $59\,{}^3\t{S}_1$ manifold.
        %\red{The weird features in the dashed lines are due to XXX.}
        The lines of different opacities show the gate infidelity for varied overall energy shift from $30\,\t{GHz}$ (most opaque) to $50\,\t{MHz}$ (least opaque), with equal logarithmic spacing. The condition used in (a) is shown in red. Note that the authors of Ref.~\cite{Ma2022} estimate their shift to be of size $\approx 26\,\t{GHz}$.
        \label{fig:rydberg-cz}
    }
\end{figure}

Although the CCCZ protocol merely extends that identified in Ref.~\cite{Levine2019} to include effects on additional states in the ququart encoding scheme, the CZ protocol requires consideration of the magnitude of the Rydberg interaction between different states in the $n\,{}^3\t{S}_1$ manifold. Specifically, we consider effects from non-uniform energy shifts on the $\ket*{m_F^{(1)}, m_F^{(2)}} \in \{ \ket{-3/2, -3/2}, \ket{+3/2, +3/2}, \ket{+3/2, -3/2}, \ket{-3/2, +3/2} \}$ two-ququart Rydberg states. We parameterize their $C_6$ values by $\eta$ as
\begin{align}
    C_6^{(-3/2, +3/2)}
        &\rightarrow \eta C_6'
    \\
    C_6^{(+3/2, -3/2)}
        &\rightarrow \eta C_6'
    \\
    C_6^{(+3/2, +3/2)}
        &\rightarrow \phantom{\eta} C_6'
    \\
    C_6^{(-3/2, -3/2)}
        &\rightarrow \phantom{\eta} C_6'
\end{align}
where $\eta = 1$ corresponds to perfect uniformity, and $\eta = 0.989$ to the result of the calculations from the previous section. Here we assume that the size of the shifts for the $\ket{\pm 3/2, \pm 3/2}$ states are equal due to symmetry, and neglect parameterization of all other pairs of Rydberg states due to selectivity in the applied pulses' detunings and polarizations. We vary $\eta$ from $-1$ to $+1$ as well as the overall magnitude of the Rydberg shift. $\eta$ affects the accumulation of phase between computational states due to off-resonant excitation. This phase shift also varies with the array spacing, and the chosen range of shift magnitudes corresponds to spacings from $\approx 2.4\,\text{$\mu$m}$ to $\approx 6.8\,\text{$\mu$m}$. We note that this issue can be avoided through the use of purely circular polarizations (rather than linear polarization perpendicular to the magnetic field) such that only the transitions shown in Fig~\ref{Fig3}(b) are driven. We show the simulated phase accrual of all two-ququart states for $\eta = 0.989$ in Fig.~\ref{fig:rydberg-cz}(a), and gate infidelity as a function of $\eta$ and overall shift magnitude in Fig.~\ref{fig:rydberg-cz}(b). In particular, we see that the gate fidelity is more robust to effects from non-uniform interactions for larger shift sizes. For expected interaction shifts of $\sim\text{GHz}$, we can tolerate an interaction disparity of $1 - \eta \gtrsim 0.95$ between the two pair-states to still obtain a gate fidelity of $\mathcal{F} \approx 0.999$.

\subsection{Simulating Entanglement Distillation and Flag-based FTQEC}
\label{Appendix:simulation_details}

In section \ref{Applications}, we proposed several possible applications related to the ququart architecture and simulated how the behavior gets improved when using ququarts. Here we discuss several details about the simulations, including error models for gates, circuit-level error models for QEC simulations, and the stabilizer lists and logical operators for the QEC code we used to perform simulations.

\subsubsection{Entanglement distillation}
\begin{figure}[t!]
    \includegraphics[width=0.8\columnwidth]{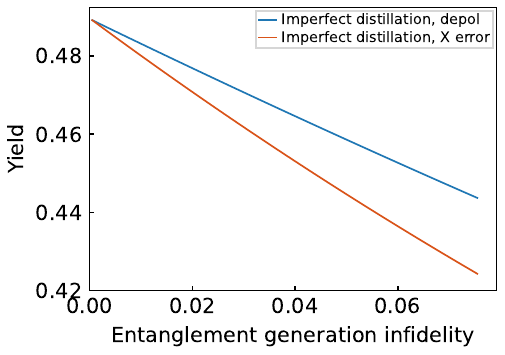}
    \caption{\label{Fig:Yield}The yield of imperfect distillation with respect to different values entanglement generation infidelity for different error models. The initial deviation from 0.5 is due to the assumed 1\% measurement error.}
\end{figure}

Figure \ref{Fig5}(a) shows the circuit of entanglement distillation, with the entanglement generation presented as two wavy lines. 
We simulate the pre- and post-distillation entanglement infidelity and yield of the circuit in Fig.~\ref{Fig5}(a), whose result is  shown in Fig.~\ref{Fig5}(b) and Fig.~\ref{Fig:Yield}. Here we fix the measurement error rate to $1\%$ and the intra-ququart gate infidelity to $0.1\%$, and vary the error rate in the two different error models to change the pre-distillation entanglement infidelity.

\subsubsection{Flag-based FTQEC}

Figure \ref{Fig6}(b) simulates how the threshold for certain codes under flag-based QEC can increase by replacing the ancilla and flag qubits with one ququart. We performed a circuit-level simulation, using the Pauli frame to track how single-qubit faults propagate through two-qubit CNOT gates during the whole circuit. The faults on physical qubits/ququarts are generated via the depolarization channel: During the state preparation each physical qubit has a probability $p$ to be replaced with a maximally mixed state, but post-selection by measuring the stabilizers ensures that the state preparation process is fault-tolerant.
The measurement has probability $p$ to be faulty, and after each inter-qubit (ququart) gate the state is replaced with a totally mixed two-qubit state with probability $p$. To show the improvement from using ququarts and due to the limitations of the Pauli frame (only uncorrelated Pauli errors on each qubit can be tracked), we treat intra-ququart CNOT gates as perfect, but existing single-qubit faults can still propagate through intra-ququart CNOT gates. After one round of circuit-level error correction, we perform another round of perfect decoding. The circuit-level decoding is denoted as successful if, after the perfect decoding, the logical state gives no logical errors.

Using the model described above, we simulate the logical error rate of $\llbracket5,1,3\rrbracket$ and Steane $\llbracket7,1,3\rrbracket$ codes using flag-based error correction scheme under different values of $p$. For convenience, we list out the stabilizers and logical operators for the $\llbracket5,1,3\rrbracket$ and $\llbracket7,1,3\rrbracket$ codes respectively:

$\llbracket5,1,3\rrbracket$:
\begin{align}
\begin{split}
    S_1 &= X_1Z_2Z_3X_4,\\
    S_2 &= X_2Z_3Z_4X_5,\\
    S_3 &= X_1X_3Z_4Z_5,\\
    S_4 &= Z_1X_2X_4Z_5, \\
    X_L &= X_1X_2X_3X_4X_5,\\
    Z_L &= Z_1Z_2Z_3Z_4Z_5.
\end{split}
\end{align}

$\llbracket7,1,3\rrbracket$:
\begin{align}
\begin{split}
    S_1 &= X_1X_2X_3X_4,\\
    S_2 &= X_2X_3X_5X_6,\\
    S_3 &= X_3X_4X_6X_7,\\
    S_4 &= Z_1Z_2Z_3Z_4,\\
    S_5 &= Z_2Z_3Z_5Z_6,\\
    S_6 &= Z_3Z_4Z_6Z_7,\\
    X_L &= X_1X_2X_3X_4X_5X_6X_7,\\
    Z_L &= Z_1Z_2Z_3Z_4Z_5Z_6Z_7.
\end{split}
\end{align}

The ququart flag shows an improved threshold, making the logical qubit more robust against the physical error rate $p$ as shown in Fig. \ref{Fig6}.

\subsection{Fusion and Fission protocol} \label{app:fusionfission}

\begin{figure}[t!]
     \centering
     \includegraphics[width=\linewidth]{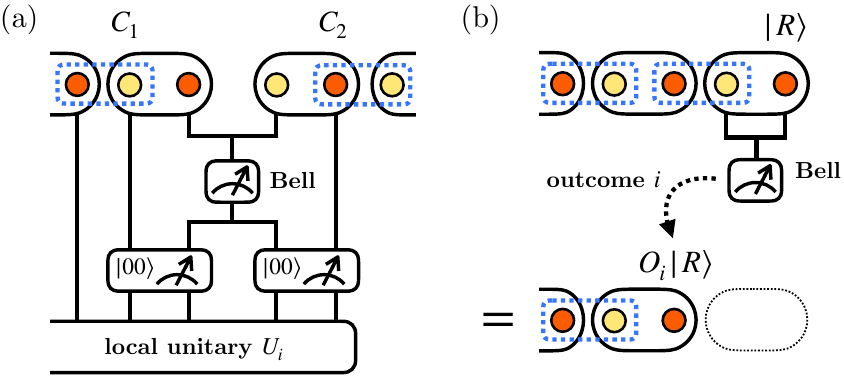}
        \caption{  
                {\bf Fusion and Fission protocols}. (a) We perform an inter-ququart Bell-basis measurement between the nuclear qubit for $C_1$ and optical qubit for $C_2$, which is followed by measurements on two ququarts that project each ququart on a singlet or triplet. Finally, we perform local unitary operations based on the Bell-basis measurement outcome $i$ (and the rearrangement based on the triplet measurement outcomes). (b) We perform a Bell-basis measurement on the last ququart. This will disentangle the last ququart from the others, leaving the remaining $L-1$ ququarts in an AKLT state. This AKLT state is as if the initial right boundary state was $O_i |R \rangle$, where $O_i$ is $X$, $Y$, or $Z$ depending on the measurement outcome.}   
        
        \label{fusionfission}
\end{figure}

In the main text, we discussed a method to realize an AKLT state. Starting from $L$ ququarts, on average we would obtain an AKLT state of length $3L/4$ with a standard deviation of $\sqrt{3L}/4$ from the Bernoulli distribution. Due to the stochastic nature of the proposed protocol, an additional step is required to achieve a specific length. There are two different approaches, bottom-up and top-down. 
 
In the bottom-up approach, we employ a \emph{fusion} process. Consider two AKLT segments \( C_1 \) and \( C_2 \) of lengths \( l_1 \) and \( l_2 \), respectively, prepared from initial spin-1/2 chains with edge states \( (L_1, R_1) \) and \( (L_2, R_2) \). These segments would form a single AKLT chain if the edge states \( R_1 \) and \( L_2 \) were in a singlet state before the triplet projection.  
This can be achieved in two steps, as illustrated in \figref{fusionfission}(a). First, we perform a singlet projection on \( (R_1, L_2) \). Second, since this singlet projection affects triplet conditions for the two ququarts to which $R_1$ and $L_2$ belong, we have to perform the triplet projection measurement again for these two ququarts. This protocol properly fuses two AKLT segments into one since this singlet projection does not affect the short-range entangled structure of other ququarts. The resulting state becomes a single AKLT chain with length $L\,{=}\,l_1 + l_2 - m$, where $m$ is the number of atoms projected onto singlet states. Note that $P(m{=}0)\,{=}\,9/16$, $P(m{=}1)\,{=}\,6/16$, and $P(m{=}2)\,{=}\,1/16$.  
  
One way to realize the singlet projection in \( (R_1, L_2) \) is by performing an inter-ququart Bell measurement on \( (R_1, L_2) \) and applying an on-site unitary gate depending on the measurement outcome. If the Bell measurement yields \( |\Psi_{S=0} \rangle := |01\rangle - |10\rangle \), we achieve a successful singlet projection. However, there are three other possible outcomes: \( X_{R_1} |\Psi_{S=0} \rangle \), \( Y_{R_1} |\Psi_{S=0} \rangle \), or \( Z_{R_1} |\Psi_{S=0} \rangle \), which are equivalent to the singlet projection up to a local unitary $O_i \in \{X,Y,Z\}$ on \( R_1 \) before the triplet projection.
We utilize the property of an SPT wavefunction: applying a symmetry transformation \( \prod_i U_i(g) \) for \( g \in G \) is equivalent to applying symmetry transformations on the left and right edge modes \( U_L(g) \) and \( U^{-1}_R(g) \), respectively~\cite{Schuch_2011}. 
Therefore, we can ``undo'' a Pauli operator on \( R_1 \) by applying a product of local unitaries \( \prod_i U_i(g) \) on the spin-1 subspace of ququarts, and the only consequence is that \( |L_1\rangle \) would be transformed into \( U_L(g) |L_1 \rangle \).
Note that $g$ for \( U_R(g)\,{\sim}\,X, Y, Z \) corresponds to $\pi$-rotation around $\hat{x}$, $\hat{y}$, and $\hat{z}$ axis respectively, and $U_i(g)$ can be easily implemented using intra-ququart gates.  
This deterministic procedure allows us to fuse two AKLT segments into a single AKLT chain, irreversibly removing information about \( R_1 \) and \( L_2 \). A similar concept has been discussed in Ref.~\onlinecite{AKLT_exp7}.
 
In the top-down approach, we employ a \emph{fission} process to reduce the AKLT state from an initial length $L'$ to the desired smaller length $L$. The idea is to convert the right-most triplet site into a singlet and remove it from the chain, as shown in \figref{fusionfission}(b). This can be achieved by performing an intra-ququart Bell measurement on optical and nuclear qubits within the right-most site.
Since the ququart is already in a state with $S=1$, the Bell basis measurement can collapse it into three possible states: \( X_1 |\Psi_{S=0} \rangle \), \( Y_1 |\Psi_{S=0} \rangle \), and \( Z_1 |\Psi_{S=0} \rangle \).
This implies that the Bell basis measurement makes it as if the last ququart was projected onto singlet up to Pauli correct at the Bell pair stage. Furthermore, this will automatically teleport the edge state $|R\rangle$ in the nuclear qubit at the site-$L'$ to the site-$(L'-1)$. 

More precisely, the right-most edge state for the AKLT segment with length $(L'-1)$ is now labeled by $O |R \rangle$ where $O$ is $X$, $Y$ or $Z$ depending on the measurement outcome. Through this fission process, we can deterministically remove each right-most ququart, resulting in a new AKLT state with a shorter length and an edge state $O_i |R\rangle$, modified according to the measurement outcome. 
We emphasize that this fission protocol (or removal protocol) can be applied in parallel for any number of ququarts. By performing intra-ququart Bell-basis measurements on the last $L'-L$ ququarts to be removed, the resulting state will be an AKLT state derived from the initial state, with the right edge mode labeled as $\prod_{i=1}^{L'-L} O_i | R \rangle$.

\section{Data Availability}
The simulation data is available from the first or corresponding author upon reasonable request.

\section{Code Availability}
The codes used to generate data for this paper are available from the first or corresponding author upon reasonable request.

\section{Acknowledgements}
We acknowledge Ken Brown, Xiao Chen, Sagar Vijay, Matteo Ippoliti, Bryan Clark, and Hannes Bernien for stimulating discussions; and Yichao Yu for critical reading of the manuscript. We acknowledge funding from the NSF QLCI for Hybrid Quantum Architectures and Networks (NSF award 2016136); the NSF PHY Division (NSF award 2112663); the NSF Quantum Interconnects Challenge for Transformational Advances in Quantum Systems (NSF award 2137642); the ONR Young Investigator Program (ONR award N00014-22-1-2311); the AFOSR Young Investigator Program (AFOSR award FA9550-23-1-0059); the Simons Investigator Award (E.A.); and the U.S. Department of Energy, Office of Science, National Quantum Information Science Research Centers.

\section{Author Contributions}
Z.J., J.Y.L., and J.P.C. conceived the main ideas of this work. Z.J. and W.H. performed the simulations. Z.J., W.H., W.K.C.S., J.Y.L., and J.P.C. prepared the figures and text for the paper with input from all authors. All authors contributed to the proofing and verification of the writing and results. J.P.C. supervised this project.

\section{Competing Interests}
The authors declare no competing interests.

\appendix

\section{Utilizing a single ququart as two effective qubits}
\label{section:Hamiltonian}
Here we describe one simple approach to control a single ququart as two effective qubits via resonant tones. 
Consider a qudit with bare Hamiltonian $H_0 = \sum_k \omega_{0,k} \dyad{k}{k}$, with an external drive given by $H_c(t)=\sum_{j,k} H_c^{jk}(t)$, where 
\begin{align} \label{eq:a}
    H_c^{jk}(t) &= \Omega_{0}^{jk} \cos({q_{jk}t+\phi_{jk}}) (\dyad{j}{k}+hc). %\nonumber \\
\end{align}

We now assume for simplicity that the presence of drives ($\Omega_{0}^{jk}\!\neq\!0 $) between states $j,k$ do not introduce parasitic light shifts to the rest of the states. 
Then, going into the rotating frame given by the bare Hamiltonian and $T_r=e^{-i H_r t}$ with $H_r=\sum_k \nu_{0,k} \dyad{k}{k}$, the effective Hamiltonian becomes
\begin{align}  
    H_\text{eff} &= T_r (H_0 + H_c(t)) T_r^\dag + H_r \nonumber \\
    &= \sum_k (\omega_{0,k}\!-\!\nu_{0,k}) \dyad{k}{k} + \sum_{jk} \frac{\Omega_0^{jk}}{2} (e^{i\theta_{jk}}\dyad{j}{k}+\text{h.c.})
\end{align}
where $\theta_{jk}=(q_{jk}-\Delta_{kj})t +\phi_{jk}$ and $\Delta_{kj}=\nu_k-\nu_j$. 
Thus applying drive frequencies $q_{jk}$ resonant with the various transition $\Delta_{kj}=\nu_k-\nu_j$, the Hamiltonian becomes time-independent.

Focusing on the ququart ($d\!=\!4$) system, we see the drives resonant with various pairs of states $\ket{j}$ yield effective two-qubit rotations with relabeling $j=\{0,1,2,3\} \mapsto \{00, 01, 10,11\}$. Below we list some pertinent drives that yield generalized CNOT's (controlled rotations or ``CRot's'') and flip-flop/SWAP Hamiltonians ($H_c^{12}$):
\begin{align}  
    H_c^{01} &\approx  \frac{\Omega_{0}^{01}}{2} \dyad{0}{0}_0 \otimes (\sigma_1^x\cos{\phi_{01}}+\sigma_1^y\sin{\phi_{01}}), \nonumber \\
    H_c^{23} &\approx  \frac{\Omega_{0}^{23}}{2} \dyad{1}{1}_0 \otimes (\sigma_1^x\cos{\phi_{23}}+\sigma_1^y\sin{\phi_{23}}), \nonumber \\
    H_c^{02} &\approx  \frac{\Omega_{0}^{02}}{2} (\sigma_0^x\cos{\phi_{02}}+\sigma_0^y\sin{\phi_{02}}) \otimes \dyad{0}{0}_1, \nonumber \\
    H_c^{13} &\approx  \frac{\Omega_{0}^{13}}{2} (\sigma_0^x\cos{\phi_{13}}+\sigma_0^y\sin{\phi_{13}}) \otimes \dyad{1}{1}_1, \nonumber \\
    H_c^{12} &\approx  \frac{\Omega_{0}^{12}}{2} (\sigma_0^x\sigma_1^x + \sigma_0^y\sigma_1^y)/2, \nonumber \\
\end{align}
where $\sigma_j$ refers to Pauli operator on the $j\text{-th}$ qubit. Note the combination of CRot's conditioned on both $\dyad{0}{0}_j$ and $\dyad{1}{1}_j$ of the $j\text{-th}$ qubit yields arbitrary single-qubit rotation on the other qubit.

Thus given both arbitrary single-qubit rotations as well as two-qubit interactions we find the use of resonant tones in a ququart, combined with state initialization and measurement, allow universal control over the effective two-qubit system. 
% We remark that not all drives are necessary for universal control, since an arbitrary rotation on a single qubit along with a two-qubit interaction should suffice.

We remark that thus far we have assumed there is no light shifts added to the system due the presence of the applied drives themselves ($\Omega_{0}^{jk}\!\neq\!0 $), which may result in, e.g., undesired effective phase evolution between various states. Such light shifts on states coupled by $H_c^{jk}$ could arise from states not only inside but also outside the $d=4$ computational subspace of interest.
Thus in practice for every choice of the $d=4$ states as well as applied controls $H_c^{jk}$ it may be necessary to identify and correct for these light shifts, for instance by specific timing or applying physical or virtual Z-gates on appropriate pairs of states via laser pulses or adjusting phases on the software end.

\section{Quantum state tomography of ququart states}
\label{section:QST}
Here we discuss two approaches to perform quantum state tomography (QST) of the target $d\!=\!4$ system, closely following earlier works~\cite{James2001, Bonk2004}. Depending on the experimentally available control, one approach may be more favorable than the other.

First, if effective single-qubit controls within the qudit (of dimension $d=2^N$) are readily accessible, it may be favorable to perform QST of the qudit as an effective $N$-qubit system, by describing the quantum state to be reconstructed $\rho$ in the Pauli basis with $(4^N\!-\!1)$ real parameters $\alpha_{ij}$. For a ququart,
\begin{align} \label{eq:measure}
    \rho &= \sum_{j,k\in\{0,x,y,z\}} \alpha_{jk}\sigma^j \otimes \sigma^k. %\nonumber \\
\end{align}
Here, $\sigma^0=\mathbb{I}$ is the identity and $\sigma^j  ~\forall~ j\in\{x,y,z\}$ are Pauli matrices. By the trace property of a density matrix $\tr{\rho}=1$, $\alpha_{00} =1/d$. Thus, to perform QST it remains to experimentally find the $(4^N-1)$ unknowns.
While there is no unique procedure to achieve QST, we describe one rather intuitive approach to identify $\rho$ via arbitrary single-qubit rotations and projective population measurements (access to projective measurement operators $P_{jk}=\dyad{jk}{jk}$, with $j,k\in\{0,1\}$).

First, notice that the direct population measurements $\tr{\rho P_{jk}} \equiv \langle P_{jk} \rangle=\langle jk| \rho |jk \rangle$ gives a system of unique equations which can be solved: 
\begin{align}  
    \langle P_{00} \rangle &= (1 + \langle\sigma^z_0 \rangle + \langle\sigma^z_1\rangle +  \langle\sigma^z_0\sigma^z_1  \rangle)/4, \nonumber \\
    \langle P_{01} \rangle &= (1 + \langle\sigma^z_0 \rangle - \langle\sigma^z_1\rangle -  \langle\sigma^z_0\sigma^z_1  \rangle)/4, \nonumber \\
    \langle P_{10} \rangle &= (1 - \langle\sigma^z_0 \rangle + \langle\sigma^z_1\rangle -  \langle\sigma^z_0\sigma^z_1  \rangle)/4, \nonumber \\
    \langle P_{11} \rangle &= (1 - \langle\sigma^z_0 \rangle - \langle\sigma^z_1\rangle +  \langle\sigma^z_0\sigma^z_1  \rangle)/4. \nonumber \\
\end{align}
Note the left-hand side (LHS) are given by experimentally measured values, while the right-hand side (RHS) is given by a linear combination of the unknowns $\alpha_{jk}$ that describe our state of interest $\rho$.

Having identified $\langle\sigma^z_j\rangle$ and $\langle\sigma^z_0\sigma^z_1  \rangle$, we can now access other Pauli strings by transforming the effective measurement operators with single-qubit rotations $R_{ab}(\theta_a,\theta_b) = R_a(\theta_a)\otimes R_b(\theta_b)$, with $a,b\in\{0,x,y\}$ and $R_a(\theta_a)=e^{-i \theta_a \sigma^a /2}$.
Concretely, by applying $R_{ab}(\theta_a,\theta_b)$ before measuring population, we will measure $P_{jk}^{ab}=R_{ab}(\theta_a,\theta_b)^\dag P_{jk} R_{ab}(\theta_a,\theta_b)$.

For instance, utilizing only $\pi/2$-pulses ($\theta_a=\theta_b =\pi/2$) for simplicity, we find
\begin{align}  
    \langle P_{00}^{xy} \rangle &= (1 - \langle\sigma^y_0 \rangle + \langle\sigma^x_1\rangle - \langle\sigma^y_0\sigma^x_1  \rangle)/4, \nonumber \\
    \langle P_{01}^{xy} \rangle &= (1 - \langle\sigma^y_0 \rangle - \langle\sigma^x_1\rangle + \langle\sigma^y_0\sigma^x_1  \rangle)/4, \nonumber \\
    \langle P_{10}^{xy} \rangle &= (1 + \langle\sigma^y_0 \rangle + \langle\sigma^x_1\rangle + \langle\sigma^y_0\sigma^x_1  \rangle)/4, \nonumber \\
    \langle P_{11}^{xy} \rangle &= (1 + \langle\sigma^y_0 \rangle - \langle\sigma^x_1\rangle - \langle\sigma^y_0\sigma^x_1  \rangle)/4. \nonumber \\
\end{align}

Therefore, with access to population measurements, $3^N=9$ unique measurement settings suffice to fully identify the $4^N-1=15$ unknowns of the (effective) $N$-qubit system and hence $\rho$. 
For a general qudit, however, all population measurements per single experiment (quantum circuit) may not be readily available either due to the physical quantum system or available controls; in such a case, up to $4^N-1$ measurements may be required to fully solve for $\rho$.
Finally, due to experimental uncertainty the reconstructed $\rho$ may not satisfy all three desired properties of density matrices; thus to remedy this, a maximum likelihood estimation (MLE) of the `true' density matrix can be reconstructed using $\rho$ as an initial estimate as pioneered by earlier works~\cite{James2001}. 

While the above approach may be favorable if effective single-qubit rotations (in the qudit) are readily accessible, the QST of the qudit as a general $d$-state system can be done with access to resonant drives between various $d$ states as pioneered in earlier works~\cite{Bonk2004}. Following~\cite{Bonk2004} we now describe the procedure to identify $\rho$ of the $d=4$ system, and for concreteness we assume access to ``single-quantum'' transitions $H_c^{01},~H_c^{12},~H_c^{23}$ as defined in section~\ref{section:Hamiltonian}.

To simplify notations in ref. \cite{Bonk2004}, we define $\rho_{jj}(U)= \tr{P_{jj} U\rho U^\dag}$, where $\rho$ is the state to be reconstructed.
Similar to above, we can start by extracting the state populations $\rho_{jj}(\mathbb{I}), ~\forall j\in \{0,1,2,3\}$.
To access the coherences $\rho_{j,k\!\neq j}$, we can now apply effective $\pi/2$-rotations (and their combinations) to $\rho$ before measurement. 
For instance, applying a $\pi/2$-pulse on the $\ket{0}\leftrightarrow\ket{1}$ transition 
along both $x$ and $y$, to call $X_{01}$ and $Y_{01}$, one can identify their coherence:
\begin{align}  
    \rho_{01} &= \Re{\rho_{01}} + i \Im{\rho_{01}} , \nonumber \\
    \Re{\rho_{01}} &= (\rho_{00}(Y_{01})-\rho_{11}(Y_{01}))/2, \nonumber \\
    \Im{\rho_{01}} &= (\rho_{00}(X_{01})-\rho_{11}(X_{01}))/2. \nonumber \\
\end{align}
Similarly, the remaining coherences $\rho_{jk}$ can be determined by applying $\pi/2$-pulses on relevant transitions:
\begin{align} 
    \Re{\rho_{12}} &= \big[\rho_{11}(Y_{12})-\rho_{22}(Y_{12})\big]/2, \nonumber \\
    \Im{\rho_{12}} &= \big[\rho_{11}(X_{12})-\rho_{22}(X_{12})\big]/2, \nonumber \\    
    \Re{\rho_{23}} &= \big[\rho_{22}(Y_{23})-\rho_{33}(Y_{23})\big]/2, \nonumber \\
    \Im{\rho_{23}} &= \big[\rho_{22}(X_{23})-\rho_{33}(X_{23})\big]/2, \nonumber \\        
    \Re{\rho_{02}} &= \big[\rho_{11}(X_{12}X_{01})-\rho_{22}(X_{12}X_{01}) \nonumber \\
        &\phantom{=}-\sqrt{2} \Im{\rho_{12}}\big]/\sqrt{2}, \nonumber \\
    \Im{\rho_{02}} &= \big[\rho_{22}(Y_{12}X_{01})-\rho_{11}(Y_{12}X_{01}) \nonumber \\
        &\phantom{=}+\sqrt{2} \Re{\rho_{12}}\big]/\sqrt{2}, \nonumber \\    
    \Re{\rho_{13}} &= \big[\rho_{22}(X_{23}X_{12})-\rho_{33}(X_{23}X_{12}) \nonumber \\
        &\phantom{=}-\sqrt{2} \Im{\rho_{23}}\big]/\sqrt{2}, \nonumber \\
    \Im{\rho_{13}} &= \big[\rho_{33}(Y_{23}X_{12})-\rho_{22}(Y_{23}X_{12}) \nonumber \\
        &\phantom{=}-\sqrt{2} \Re{\rho_{23}}\big]/\sqrt{2}, \nonumber \\
    \Re{\rho_{03}} &= \rho_{22}(Y_{23}Y_{12}Y_{01})-\rho_{33}(Y_{23}Y_{12}Y_{01}) \nonumber \\
        &\phantom{=}-\sqrt{2} \Re{\rho_{23}} + \Re{\rho_{13}}, \nonumber \\
    \Im{\rho_{03}} &= \rho_{33}(X_{23}X_{12}X_{01})-\rho_{22}(X_{23}X_{12}X_{01}) \nonumber \\
        &\phantom{=}+\sqrt{2} \Im{\rho_{23}} + \Re{\rho_{13}}. \nonumber \\
\end{align}
As noted above, the experimentally reconstructed $\rho$ could be further improved via MLE~\cite{James2001}.

\section{Intra-ququart gates}
\label{section:intra-ququart-gates}
Here we analyze the intra-ququart gates shown in Fig.~\ref{Fig2}. We note that there are two operating regimes: 
\noindent
1) Low magnetic field ($B\lesssim5$ G) where the nuclear spin splitting is $\sim$kHz-level and thus negligible compared to the Rabi frequencies used for clock, stimulated Raman, and Rydberg transitions. This regime was used in Refs.~\cite{Barnes2021,Ma2022,Jenkins2022,Ma2023}. 

\noindent
2) High magnetic field ($B\gtrsim100$ G) where the differential nuclear spin splitting between the ground and metastable manifolds is non-negligible. This regime was used in Refs.~\cite{Huie2023,Norcia2023} for QND readout of the ground nuclear spin. Our proposed ququart readout techniques rely on this ability. However, all other operations could be performed in the low-field regime.

In this Appendix, we encounter the need for several light-shifting operations. We emphasize that many of these light shifts could be viewed as virtual $\sigma_z$ gates that could be addressed more efficiently at the circuit level since they are deterministic and trackable. In some cases, only global light shift beams may be required to address these ququart-related shifts. In full generality, our ququart architecture will require local gate operations and local light shift operations. However, we believe that this issue is largely also present for qubit-based architectures. Particularly for `omg' and dual-species approaches, two sets of gate operations may be required at both the global and local levels. For single-species alkali approaches, several `zones' may be required for various combinations of local and semi-local operations.

\subsection{`o'-qubit X rotations}
\label{o-qubit-X}

\begin{figure}[t]
    \includegraphics[width=\columnwidth]{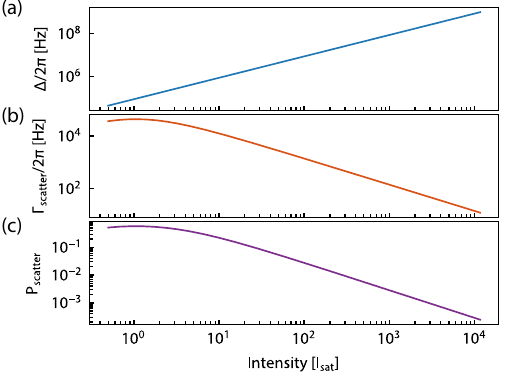}
    \caption{
        \textbf{Equalization of nuclear splittings via light shift}. (a) The detuning to the red of the ${}^1\t{S}_0 ~ \ket{m_F = +1/2} \leftrightarrow {}^3\t{P}_1 ~ F = 1/2 ~ \ket{m_F = -1/2}$ transition required for the total energy differences between the ground states in the $^{1}\t{S}_0$ and ${}^3\t{P}_0$ manifolds to be equal, as a function of applied laser intensity in units of the saturation intensity, $I_\t{sat} \approx 0.14\,\t{mW}/\t{cm}^2$. The detuning is held fixed to this function in (b) and (c). (b) The effective scattering rate on this transition while light is applied. (c) The total probability of scattering a photon for total illumination time $20\,\mu\t{s}$, corresponding to a single ``o''-qubit X rotation.
        \label{fig:nuclear-split-shift}
    }
\end{figure}

A pure ``o''-qubit X operation requires that the state rotation be completely agnostic to the ``n''-sector qubit. To facilitate ``o''-qubit X rotations as shown in Fig.~\ref{Fig2}(a), we propose two approaches. The first is to operate in the low-field regime for which the nuclear spin degree of freedom has been demonstrated to be preserved during optical clock transitions~\cite{Lis2023}. The second approach is designed for high-field operation. We propose to apply a light shift such that the total energy difference between the $\ket{00}$ and $\ket{01}$ states becomes equal to that between $\ket{10}$ and $\ket{11}$. With either approach, both transitions shown in Fig.~\ref{Fig2}(a) are driven by the same, monochromatic pulse. Hence, driving an ``o''-qubit X rotation is tantamount to driving two identical two-level systems in parallel, thus making it additionally compatible with various other methods to combine high-fidelity population transfer with preservation of the atom's motional state in the trap~\cite{Lis2023, Cummins2003}.

We now briefly discuss both the single-state light shift and the motion-preserving ``o'' gates.

%Since the frequencies of the $\ket{00} \leftrightarrow \ket{10}$ and $\ket{01} \leftrightarrow \ket{11}$ transitions then become equal under this condition as well, driving an ``o''-qubit X rotation also becomes equivalent to driving only one of these transitions, thus making it additionally compatible with various other methods to combine high-fidelity population transfer with preservation of the ququart's motional state in a trapping apparatus \cite{Lis2023, Cummins2003}.

\subsubsection{Single-state light shifts}
\label{section:light_shift}
In the presence of a $120\,\t{G}$ magnetic field, the Zeeman splitting between $\ket{10}$ and $\ket{11}$ is $\approx50\,\t{kHz}$ larger than that between $\ket{00}$ and $\ket{01}$. Perhaps the most convenient way to accomplish the desired light shift is by applying {global} red-detuned light with $\sigma^-$ polarization, far-detuned from the $\ket{01} \leftrightarrow {}^3\t{P}_1 ~ F = 1/2 ~ \ket{m_F = -1/2}$ transition. The detuning required for the ground nuclear spin splitting to match that of the metastable state is shown in Fig.~\ref{fig:nuclear-split-shift}(a) as a function of the intensity of the applied light. Fig.~\ref{fig:nuclear-split-shift}(b) and (c) show the scattering rate and probability of scattering for an assumed application time of $\approx 20\,\mu\t{s}$ corresponding to a single ``o''-qubit gate (see below). We require an intensity of $I / I_\t{sat} \approx 2.8 \times 10^3$ and detuning $\Delta / 2 \pi \approx 240\,\text{MHz}$ for a scattering probability of $\lesssim 10^{-3}$.

%time, respectively. We consider two scenarios: If we apply the light shift only during the ``o''-qubit gate for gate time $\approx20$ $\mu$s (see below), we require an intensity of $I / I_\t{sat} \approx 2.8 \times 10^3$ and detuning $\Delta/2\pi\approx 240\,\text{MHz}$ for a scattering probability of $\lesssim 10^{-3}$. If we apply the light shift during the entire circuit, generically assumed to be $\approx2$ ms when neglecting mid-circuit measurements, we require an intensity of $I / I_\t{sat} \approx 1.7 \times 10^6$ and detuning $\Delta/2\pi\approx 140\,\text{GHz}$ for a scattering probability of $\lesssim 10^{-3}$. The former case can be realized with a global beam of reasonable power while the latter case is probably unrealistic.

%Under this condition, we find that the associated probability of scattering a photon on this transition is at the $\lesssim 10^{-3}$ for $I / I_\t{sat} \gtrsim 100$ over $\sim 100\,\t{ns}$ time scales (representing application of the shift while driving the ``o''-qubit rotation). Alternatively, one can apply $I / I_\t{sat} \gtrsim 10^4$ intensities with $\sim 10\,\t{GHz}$-scale detunings over $\sim 100\,\mu\t{s}$ (e.g. applying the shift for an entire circuit) with approximately equal scattering probability instead [see Fig.~\ref{fig:nuclear-split-shift}(c)].

\subsubsection{Motional state preserving ``o''-qubit rotations}
% \red{[\dots]}

Under a magic trapping condition, one challenge towards achieving high-fidelity operations on the ``o'' qubit is eliminating the effects from motional states. Even after the atoms are cooled to motional ground state, when the optical Rabi frequency is comparable to or larger than the trap frequency (high-Rabi regime), motional sideband transitions are off-resonantly excited during clock operations, leading to leakage from qubit (ququart) states to other motional states. This issue can be solved by operating at Rabi frequency much smaller than the trap frequency, at the cost of longer gate time. Alternative faster approaches via composite pulse sequences have been demonstrated for ground-clock shelving, which in ququart language represents ``o''-qubit $\pi$-rotations~\cite{Lis2023}. Here we present two motional-state preserving ``o''-qubit gate designs for $\pi$-rotations and Hadamard gates, both working at high-Rabi regime. 
% Compared with the solution in ref.~\cite{Lis2023}, our $\pi$-rotation pulse design requires less time under the same Rabi frequency and 

\begin{figure}[t]
\includegraphics[width=\columnwidth]{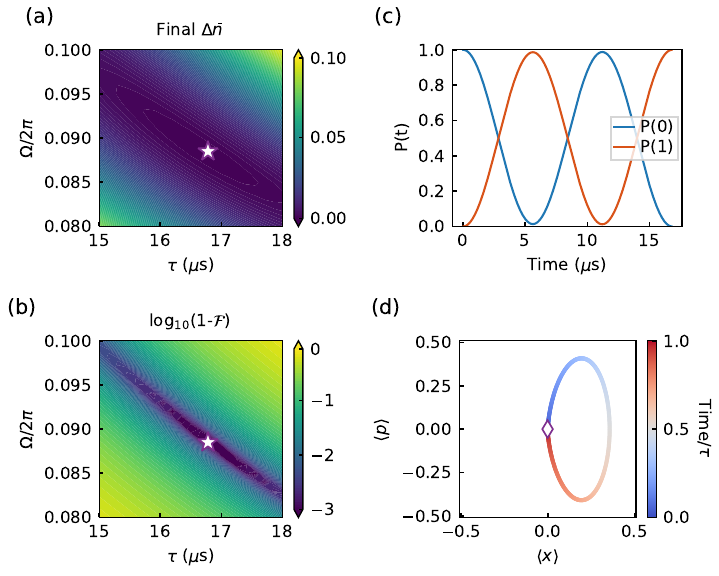}
\caption{\label{fig:O-qubit_pi}
 {\bf Clock motion-insensitive $\pi$-rotation.} (a)(b) Theoretical $\Delta\bar{n}$ and gate infidelity under different gate time and Rabi frequency, with trap frequency fixed to 30 kHz. The $\Delta\bar{n}$ and infidelity are calculated by starting from a product of qubit state $\ket{0}$ and thermal ground state. We see that there exists a region with minimal change in $\bar{n}$ and high final state fidelity. The stars indicate the conditions under which Rabi oscillations are shown in (c) and (d). (c) Spin evolution under the parameters labeled in (a) ($\Omega/2\pi=88.5$ kHz, gate time 16.78 $\mu$s). At the end of gate time the two spins are flipped. (d) Simulation of the phase space trajectory of the motional state during the evolution. The diamond denotes the motional state at the end of the gate.}
\end{figure}

Instead of using composite pulse sequences, we apply a continuous, modulation-free clock pulse on resonance to the clock carrier transition, with total gate time approximately 3 times the clock $\pi$-time, as shown in Fig.~\ref{fig:O-qubit_pi}(b). Figure \ref{fig:O-qubit_pi}(a) shows the theoretical change of the average motional quanta after the pulse and gate infidelity over certain range of parameter space, assuming the atoms start from motional ground state. We find that with a certain ratio of gate time $\tau$, Rabi frequency $\Omega$ and trap frequency $\omega$ ($\Omega\approx 3\omega$, $\tau\approx 3\pi/\Omega$), we can achieve a $\sim 99.9\%$ fidelity $\pi$-pulse while preserving the motional ground state, labeled with a star in Fig.~\ref{fig:O-qubit_pi}. Figure~\ref{fig:O-qubit_pi}(c) shows the evolution of the motional wave packet in $\langle x \rangle$-$\langle p\rangle$ phase space. The trajectory returns back to the starting point, indicating the motional state remains unchanged together with Fig.~\ref{fig:O-qubit_pi}(a).

\begin{figure}[t]
\includegraphics[width=\columnwidth]{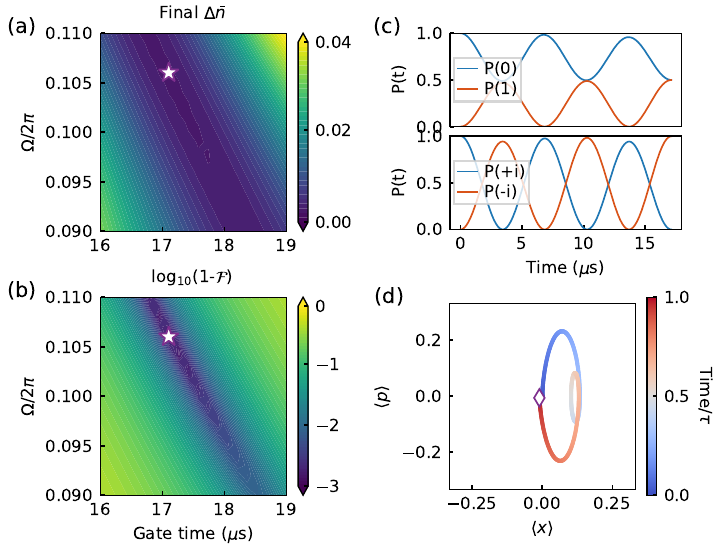}
\caption{\label{fig:O-qubit_Hadamard}
 {\bf Clock motion-insensitive Hadamard gate.} 
 (a)(b) Theoretical $\Delta\bar{n}$ and gate infidelity under different gate time and Rabi frequency, with trap frequency fixed to 30 kHz and starting from motional ground state. The stars indicate the conditions under which Rabi oscillations are shown in (c) and (d). (c) Spin evolution under the parameters labeled by the star ($\omega/2\pi=30$ kHz, $\delta/2\pi = 100$ kHz, $\Omega/2\pi = 106$ kHz, gate time 17.1 $\mu$s) starting from $\ket{0}$ and $\ket{+i}$ states respectively. At the end of gate time, $\ket{0}$ state transforms to $\ket{+}$ state and $\ket{+i}$ state transforms to $\ket{-i}$. (d) Simulation of the phase space trajectory of the motional state during the evolution, starting from $\ket{0}$ spin state. The diamond denotes the motional state at the end of the gate.
 % ( $\omega/2\pi=30$ kHz, $\delta/2\pi = 100$ kHz, $\Omega/2\pi = 106$ kHz, gate time 17.1 $\mu$s, starting from motional ground state)
 }
\end{figure}

Realizing arbitrary-angle rotations while preserving motional states is, however, non-trivial. To overcome this and to move towards a motional state-preserving universal gate set on the ``o''-qubit, we instead use a detuned clock Rabi oscillation to perform a Hadamard gate on the ``o'' qubit. Figure~\ref{fig:O-qubit_Hadamard} shows the change of average motional quanta, gate infidelity, pulse design and evolution in phase of the Hadamard gate. Similar to the $\pi$-rotation, the Hadamard pulse is also modulation-free, as is shown in Fig.~\ref{fig:O-qubit_Hadamard}(b). After searching through the parameter space, we find a set of optimized parameter setting (detuning $\delta\approx\Omega\sim 3.4\omega$, $\tau = 3\pi/\Omega$), labeled with a star in Fig.~\ref{fig:O-qubit_Hadamard}(a). The evolution in phase space also returns to the starting point as is shown in Fig.~\ref{fig:O-qubit_Hadamard}(c).

The above gate schemes only consider a two-level system. To act as an ``o'' qubit gate in a ququart, we first use the methods described in section~\ref{section:light_shift} to add single-state light shift to one of the ququart state, such that the energy difference between $\ket{00}$ and $\ket{10}$ and between $\ket{01}$ and $\ket{11}$ are the same. Then one $\pi$-polarized clock beam can be used to drive both pairs of clock rotations simultaneously while being agnostic to the ``n'' qubit.

% X rotations on the ``o''-sector qubit were simulated numerically with motional effects under a magic-trapping condition. Specifically, we work in the ququart-motional Fock basis with states of the form $\ket{ab, n} \equiv \ket{ab} \otimes \ket{n}$ where $a, b \in \{0, 1\}$ for ququart states and $n \in \mathbb{N}_0$ is the motional index and simulate dynamics due to interaction with light tuned close to the $578\,\text{nm}$ optical clock transition.

% Numerical simulations with motional effects; calculate the fidelity of a $\pi$-pulse, phase accrual in the nuclear sector, qubit state purity after the pulse, etc.

\subsection{``n''-qubit X rotations}
X rotations on the ``n'' qubit are driven via two stimulated Raman process applied simultaneously to the ground and clock manifolds. As shown in Fig.~\ref{Fig2}(b), the two clock states will use the ${}^3\t{D}_1 ~ F = 1/2 ~ m_F = -1/2$ state as the intermediate in the process, while the ground states will use the analogous state in the ${}^3\t{P}_1$ manifold. Although it is also possible to drive the two appropriate magnetic transitions simultaneously with an oscillating magnetic field when the ground and clock manifolds have equalized splittings (see the previous section), we prefer this scheme due to its access to $\sim \t{MHz}$-scale Rabi frequencies~\cite{Chen2022}.

\begin{figure}[t]
\includegraphics[width=\columnwidth]{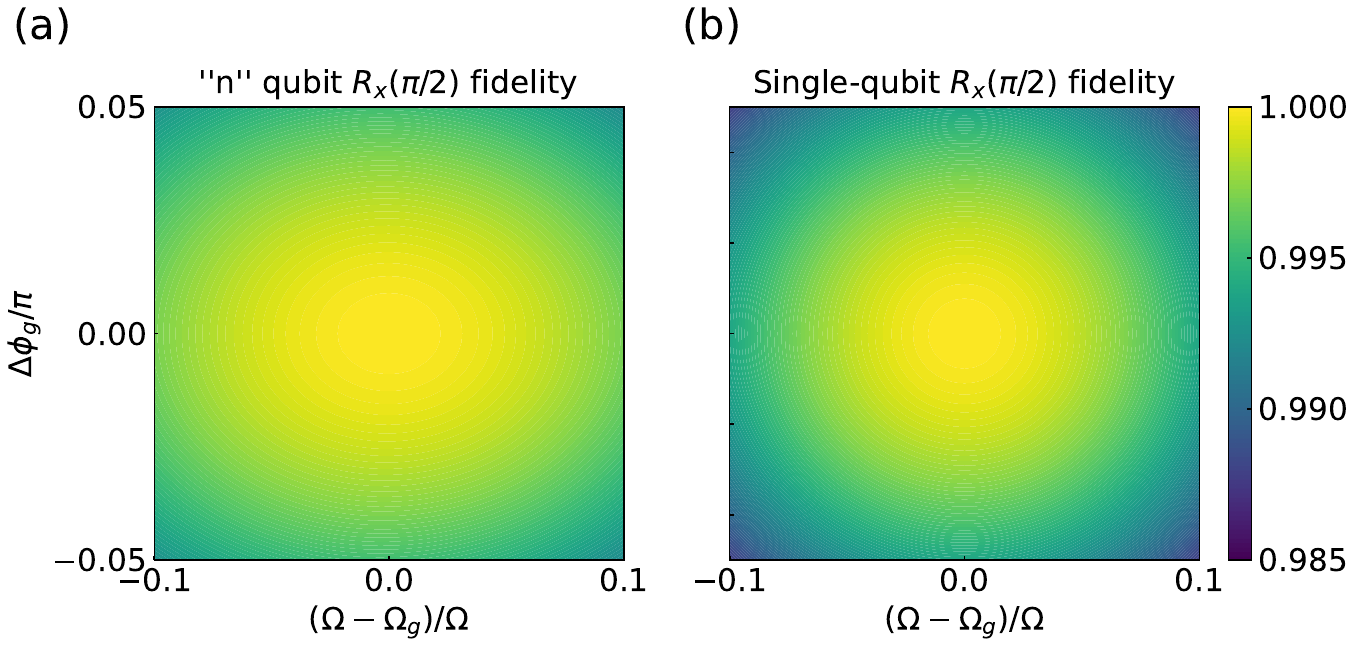}
\caption{\label{fig:n-fidelity}
 {\bf ``n'' qubit rotation fidelity under mismatching between $^1$S$_0$ and $^3$P$_0$ Raman rotations compared to single-qubit rotations with non-perfect parameters.} (a) We fixed the $^3$P$_0$ Raman transition phase and Rabi frequency to that of a perfect $R_X(\pi/2)$ rotation, and varied the phase offset $\Delta\phi$ and Rabi frequency offset $\Omega_g$ from ideal setting $\Omega$ of $^1$S$_0$ Raman transition to extract the ``n'' qubit X rotation fidelity, such that when $\Delta_\phi=0$ and $\Omega_g=\Omega$ we yield a perfect ``n'' qubit $R_X(\pi/2)$ rotation. With $-0.1 < (\Omega_g - \Omega)/\Omega < 0.1$ and $-\pi/20 < \Delta\phi_g< \pi/20$, we observe a ``n'' qubit $R_X(\pi/2)$ fidelity of $>99\%$. (b) Under the same phase and Rabi frequency offset, single-qubit rotations perform lower fidelity.}
\end{figure}

More precisely, we consider for both Raman processes illumination of the ququart by two beams: one with $\sigma^-$ polarization and one with $\pi$ polarization. Using single-beam Rabi frequencies of $\sim 40\,\t{MHz}$ with detunings of $\approx 800\,\t{MHz}$ to the excited states, it is possible to tune the pulse timings such that the accrued phase on the ``o'' qubit during the ``n'' qubit rotation due to light shifts from the Raman beams is zero modulo $2 \pi$, while additionally maintaining effective Raman Rabi frequencies of $\approx 1\,\t{MHz}$ and an idealized $\pi$-rotation infidelity of $\lesssim 10^{-3}$. We note that Raman beam intensity fluctuations will introduce fluctuations in the phase shift of the ``o'' qubit, but such fluctuations will similarly affect the ability to perform high-fidelity Raman rotations. Thus, we believe that no additional fidelity constraints are imposed on the Raman ``n''-qubit rotations due to the addition of the ``o''-qubit in the computational space. 
In addition, we simulate how a mismatch between the Raman rotations in the $^1$S$_0$ and $^3$P$_0$ manifolds can affect the effective ``n'' qubit rotations. Figure \ref{fig:n-fidelity} shows the ``n'' qubit $R_X(\pi/2)$ fidelity under mismatched rotations compared with non-ideal single qubit $R_X(\pi/2)$ rotations. We see that the ququart case [Fig.~\ref{fig:n-fidelity}(a)] is not worse than its qubit-based counterpart [Fig.~\ref{fig:n-fidelity}(b)].
% In practice, when driving simultaneous Raman transitions in $^1$S$_0$ and $^3$P$_0$ manifolds, mismatch between Raman transition parameters effectively entangles ``o'' and ``n'' qubits and decreases ``n'' qubit rotation fidelity. We studied how much Rabi frequency and phase mismatch affects the ``n'' qubit $\pi/2$-rotation fidelity as is shown in Fig. \ref{fig:n-fidelity}.

% \red{Jake: not entirely true. Emphasize that it is ok if the Rabi frequencies are not the same}

\subsection{``o''- and ``n''-qubit Z rotations}
\label{o-n-qubit-Z}
Z rotations on the ``o'' and ``n'' sectors are to be accomplished by selectively applying light shifts. Specifically, we seek to apply a uniform light shift to the two ground ququart states for an ``o''-qubit Z rotation and a uniform light shift to the two $m_F = +1/2$ ququart states for a ``n''-qubit Z rotation. Incorporating the beam described in section~\ref{o-qubit-X}, the ``n''-qubit light shift can be applied using that same beam, as well as another tuned analogously to the ${}^3\t{D}_1 ~ F = 1/2$ manifold with the same $\sigma^-$ polarization. The ``o''-qubit light shift could be applied with a linearly polarized, far-detuned laser since the goal is to essentially produce a non-`clock-magic' optical potential. Specifically, a wavelength of 532 nm has a large differential scalar polarizability for the ${}^1\t{S}_0$ and ${}^3\t{P}_0$ states. This approach was used to light-shift atoms out of resonance with the clock laser for mid-circuit operations~\cite{Lis2023}. To achieve individually-addressed ``o''- and ``n''-qubit Z rotations, one can either use tightly-focused beams \cite{Lis2023} or shuttle atoms between zones for different manipulations \cite{Bluvstein2022}.

%The ``o''-qubit light shift can be applied using a $\pi$-polarized beam detuned from the same ${}^3\t{P}_1$ manifold, likely also requiring use of the $\sigma^-$-polarized ${}^3\t{P}_1$ beam discussed in Appendix~\ref{o-qubit-X}, due to a $\approx 336\,\t{MHz}$ Zeeman splitting in the excited state manifold at $120\,\t{G}$.

\subsection{Intra-ququart CNOT and SWAP, and compensatory Z rotations}
During application of both the intra-ququart CNOT and SWAP gates [see Fig.~\ref{Fig2}(c) and \ref{Fig2}(d)], two of the ququart states accrue phases relative to the others due to light shifts. In the CNOT case $\ket{10}$ and $\ket{11}$ are shifted due to the operation itself, while in the SWAP case $\ket{00}$ and $\ket{11}$ are shifted due to various other transitions close to $578\,\t{nm}$, primarily that from the ground states to the ${}^3\t{P}_1$ manifold at $556\,\t{nm}$. Both cases amount to a combination of a Z rotation on the ``o'' qubit with a differential Z rotation between the ground and metastable nuclear spin qubits, and can hence be compensated via applied {local} light shifts. For instance, the $\sigma^-$-polarized beam detuned from the ${}^3\t{P}_1 ~ F = 1/2$ manifold from section~\ref{o-qubit-X} can be used to create a shift of the ground nuclear spin splitting. This can be combined with the far off-resonant non-`clock-magic' optical trap from section~\ref{o-n-qubit-Z} that creates a shift between the ground and metastable manifolds.

\section{Readout}
\label{section:readout}
\subsection{The effect of imperfections on the two-shot ququart readout} 

\begin{figure*}[t]
    \includegraphics[width=\textwidth]{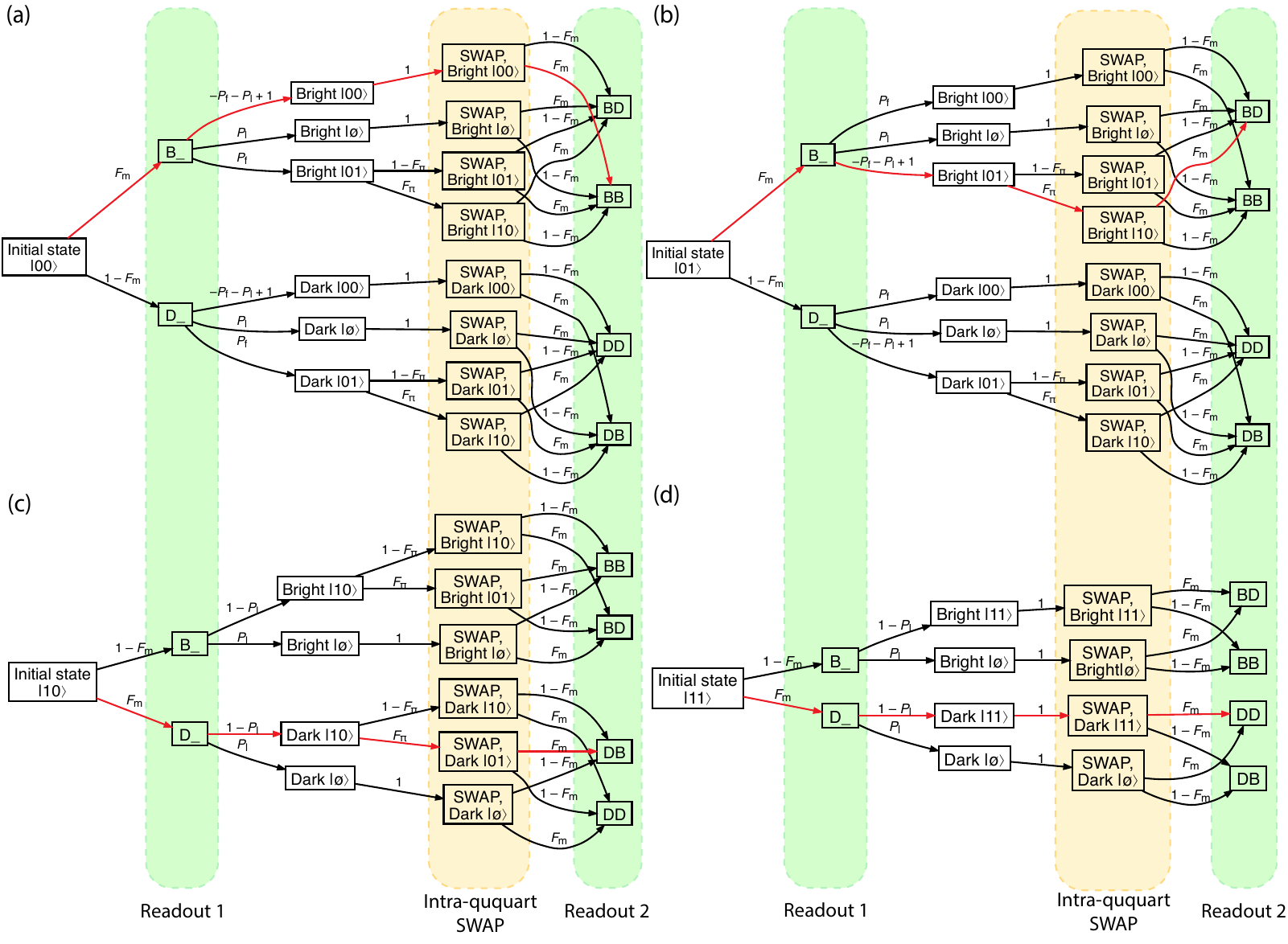}
    \caption{\label{Fig:ReadoutInfid}The effects of readout infidelity, atom loss, ground state flipping, and intra-ququart SWAP fidelity on the 2-shot readout result for (a) $\ket{00}$, (b) $\ket{01}$, (c) $\ket{10}$ and (d) $\ket{01}$ initial states. The red arrows denote the error-free path to the expected result for each initial state.}
\end{figure*}

The two-shot ququart state readout is based on the repetitive QND readout protocol of nuclear ground state qubit from ref. \cite{Huie2023}. Compared with single-shot readout, more sources of imperfections can affect the two-shot readout fidelity: (a) The single-shot readout fidelity $F_m$, (b) the atom loss probability after first readout $P_l$, (c) the ground state flip rate $P_f$, if the atom scatters photons during the first readout, and (d) the intra-ququart SWAP fidelity $F_{\pi}$. Figure~\ref{Fig:ReadoutInfid} shows how these imperfections can cause different two-shot measurement results for different initial states. The dark/bright result from each readout is indicated by the notation D/B. The red arrows indicate the error-free path, while errors occurring at different steps can lead to other wrong readout results.
% Due to the mechanism of readout circuit, the readout error is biased: $\ket{11}$ state has less branches in the graph therefore has less probability to give wrong readout result. 
For instance, if we take $F_m = F_{\pi} = 99.9\%$, $P_f = 0.1\%$ and $P_{l} = 1\%$, the probabilities of returning the right result conditioned on certain input states are
\begin{align}
\begin{split}
    P(BB|\ket{00}) = 98.7\%,\\
    P(BD|\ket{01}) = 99.6\%,\\
    P(DB|\ket{10}) = 98.7\%,\\
    P(DD|\ket{11}) = 99.8\%.
\end{split}
\end{align}

\subsection{Light shift of $|01\rangle$ due to $|00\rangle$ readout}
\label{section:LSReadout}
During single-state readout [see Fig.~\ref{Fig4}(b)], the $\ket{01}$ ground state incurs a light shift. Referring to previous experimental work~\cite{Huie2023} for QND probing conditions, we set the intensities of the probe light to $I \approx I_\t{sat}$ on all polarizations and the detuning from the targeted stretched transition to $\Delta \approx \Gamma$. In this configuration, the probe light produces a light shift of $\approx 30\,\t{Hz}$ on $\ket{01}$. For a $\tau \approx 10\,\t{ms}$ probe time, this results in the accumulation of a $\approx 1.9\,\t{rad}$ phase, which must be compensated with either an additional blue-detuned tone in the probing light or an applied light shift using the {global} beam described in section~\ref{o-qubit-X}. It may also be possible to choose other probe light parameters or timing such that the accrued phase is exactly $2\pi$. In any case, it will be important that the probe intensity and detuning are well controlled -- beyond the level typically required for readout.

\bibliography{library}

\clearpage

\end{document}